\documentclass[fleqn,usenatbib]{mnras}

\usepackage[utf8]{inputenc}

\usepackage{acronym}
\usepackage{ae,aecompl}
\usepackage{booktabs}
\usepackage{dcolumn}
\usepackage{graphicx}
\usepackage{amsmath,amsfonts,amssymb}
\usepackage{mathrsfs}
\usepackage[normalem]{ulem}
\usepackage{epstopdf}

\usepackage[T1]{fontenc}

\let\oldhref\href
\renewcommand{\href}[2]{\oldhref{#1}{\hbox{#2}}}

\title[CCSN neutrinos by long-term 2D simulations]{Supernova neutrino signals based on long-term axisymmetric simulations}

\author[H. Nagakura et al.]{
Hiroki Nagakura$^{1}$\thanks{E-mail: hirokin@astro.princeton.edu},
Adam Burrows$^{1}$,
David Vartanyan$^{2}$
\\
$^{1}$Department of Astrophysical Sciences, Princeton University, 4 Ivy Lane, Princeton, NJ 08544, USA\\
$^{2}$Astronomy Department and Theoretical Astrophysics Center, University of California, Berkeley, CA 94720, USA\\
}

\date{Accepted XXX. Received YYY; in original form ZZZ}

\pubyear{2020}

\begin{document}
\label{firstpage}
\pagerange{\pageref{firstpage}--\pageref{lastpage}}
\maketitle

\begin{abstract}
We study theoretical neutrino signals from core-collapse supernova (CCSN) computed using axisymmetric CCSN simulations that cover the post-bounce phase up to $\sim 4$~s. We provide basic quantities of the neutrino signals such as event rates, energy spectra, and cumulative number of events at some terrestrial neutrino detectors, and then discuss some new features in the late phase that emerge in our models. Contrary to popular belief, neutrino emissions in the late phase are not always steady, but rather have temporal fluctuations, the vigor of which hinges on the CCSN model and neutrino flavor. We find that such temporal variations are not primarily driven by proto-neutron star (PNS) convection, but by fallback accretion in exploding models. We assess the detectability of these temporal variations, and find that IceCube is the most promising detector with which to resolve them. We also update fitting formulae first proposed in our previous paper for which the total neutrino energy (TONE) emitted at the CCSN source is estimated from the cumulative number of events in each detector. This will be a powerful technique with which to analyze real observations, particularly for low-statistics data.
\end{abstract}


\section{Introduction}\label{sec:intro} 
Core-collapse supernovae (CCSNe) are catastrophic explosions of massive stars ($\gtrsim 8 M_{\sun}$) and cosmic factories of neutron stars (NSs) and black holes (BHs). The physical state of a NS has been a mystery since the discovery of pulsars \citep{1968Natur.217..709H}. There remains some ambiguity in the microscopic properties of neutron-star matter and in the properties of measured NSs \citep[see, e.g.,][]{2016ARA&A..54..401O}. The formation processes during CCSN explosions seem to account for some of this diversity. The next nearby CCSN is expected to provide via neutrinos and gravitational waves detailed information on the dynamics of the explosion mechanism and on NS (or BH) formation. This has motivated a multi-decade effort to develop realistic theoretical models. However, large-scale numerical simulations of CCSN are required that incorporate multi-scale and multi-physics processes. These simulations must cover the long-term post-bounce evolution for a wide range of progenitors in order to develop a comprehensive understanding of CCSN dynamics and of the formation process of compact remnants. This is a grand challenge in computational astrophysics and motivates this paper.

Theoretically, there are indications that fluid dynamics, nucleosynthesis, and neutrinos/gravitational-wave emission, key ingredients in CCSN physics, depend strongly upon dimension, except for the lightest progenitors (perhaps $\lesssim 10 M_{\sun}$). This insight is a byprodcut of the remarkable progress in multi-dimensional (multi-D) CCSN modeling during the last decades. Three-dimensional (3D) models \citep[see, e.g., ][]{2015ApJ...807L..31L,2016ApJ...831...98R,2018ApJ...865...81O,2018MNRAS.477L..80K,2019MNRAS.482..351V,2019MNRAS.490.4622N,2019MNRAS.484.3307M,2019ApJ...873...45G,2019PhRvD.100f3018W,2020MNRAS.491.2715B,2020MNRAS.492.5764N,2020ApJ...903...82I,2020arXiv201002453P,2020arXiv201010506B} are now available incorporating different implementations of the input physics. In addition, great progress has been made in axisymmetric (2D) models with full Boltzmann (multi-energy, multi-angle and multi-species) neutrino transport \citep{2018ApJ...854..136N,2019ApJ...880L..28N,2020ApJ...902..150H}, covering various types of progenitors \citep{2015PASJ...67..107N,2016ApJ...825....6S,2021Natur.589...29B}, and in long-term (> 1s post-bounce) simulation \citep{2016MNRAS.461.3296N,2021Natur.589...29B}. Although 2D and 3D CCSN models differ in the fidelity with which they treat turbulence and convection, 2D models are much more realistic than those in spherical symmetry. We have observed that they show similar explodability and neutrino emission characteristics to those in 3D \citep[see, e.g.,][]{2019MNRAS.482..351V,2019MNRAS.490.4622N}\footnote{But see also \citep{2012ApJ...755..138H,2012ApJ...759....5B,2013ApJ...775...35C,2013ApJ...765..123N,2014ApJ...786...83T} for a discussion of the differences in explodability between 2D and 3D models.}. This motivates us to use long-term 2D models to study aspects of CCSN physics when corresponding 3D models are not yet available.

Recently, we conducted a comprehensive study of the neutrino signals of our 3D CCSN models \citep{2021MNRAS.500..696N} in some representative terrestrial detectors. However, this previous study was limited to the early post-bounce phase ($\lesssim 1$~s) due to the computational expense of longer-term 3D simulations. This is an obvious limitation, since most of the neutrinos are emitted after $1$~s \citep[see the case in 1987A,][]{1987PhRvL..58.1490H,1987PhRvL..58.1494B}. Furthermore, the detailed analysis of the NS formation process using neutrino signals requires theoretical models that covers this late phase. We suggest that shorter-term 3D models can profitably be complemented by longer-term 2D models, since the latter require significantly fewer computational resources. It should be mentioned that we find that the time evolution of angle-averaged neutrino signals for the 2D and 3D models are very similar in the early phase, and this suggests that 2D models manifest essentially the same characteristics as 3D models during the late phase as well.

Much effort have been already expended in the theoretical study of neutrino signals covering later post-bounce phases \citep{2010PhRvL.104y1101H,2012PhRvL.108f1103R,2019ApJ...881..139S,2020ApJ...898..139W,2020arXiv200807070S,2020arXiv200804340W,2021PTEP.2021b3E01M}. In these studies, however, the neutrino signals have been computed by employing either toy models or spherically symmetric simulations employing artificial prescriptions. For instance, the timing of shock revival is controlled by hand and the subsequent evolution of the system has been treated with crude approximations. These simplifications may ignore important characteristics in neutrino signals and also smear out  progenitor dependent features. The lack of PNS convection in spherically symmetric models is another concern, although the convection has been effectively treated by mixing-length theory \citep[see, e.g.,][]{2012PhRvL.108f1103R}. We are still, however, far from fully understanding the detailed properties of the PNS convection, indicating that the robust conclusion of the role of convection for neutrino emissions is still missing. Direct hydrodynamical modeling of PNS convection is, hence, highly preferred in order to study the impact of PNS convection on neutrino signals appropriately.

In this paper, we analyze neutrino signals computed for our recent 2D CCSN simulations that cover the later phase ($\sim 4$~s) and a wide range of progenitors. These CCSN models contain both explosions/non-explosion cases. It should be mentioned that the interior of the PNS is not excised in the simulations, indicating that our CCSN models include all hydrodynamical feedback to the neutrino signals in a self-consistent manner. These high-fidelity numerical models reveal a rich diversity of neutrino signals across the progenitor continuum. We provide some useful fitting formulae which can be directly applied to real observations to estimate the total neutrino energy (TONE) emitted at the CCSN source from purely observed quantities. We apply our method to the neutrino data from SN 1987A. By combining other recent observational constraints regarding the NS equation-of-state (EOS), we place a constraint on the mass of a NS in SN1987A. It should be mentioned that our neutrino data are publicly available\footnote{\url{https://www.astro.princeton.edu/~burrows/nu-emissions.2d}}. These data will prove useful for more detailed detector simulations and to develop new methods and pipelines with which to analyze neutrino signals.

This paper is organized as follows. We first describe some essential aspects of our methods and models in Sec.~\ref{sec:methodsandmodels}. Sec.~\ref{sec:results} contains the bulk of this paper, in which all the results are described. Finally, we conclude this paper with a summary and discussion in Sec.~\ref{sec:sumconc}.

\section{Methods and models}\label{sec:methodsandmodels} 
\subsection{Axisymmetric CCSN models}\label{sec:axisymmetricCCSN}
First, we briefly summarize our 2D CCSN models. The simulations were carried out using our neutrino-radiation hydrodynamic code F{\sc{ornax}}, designed to capture realistic multi-D features of the dynamics by incorporating up-to-date input physics and numerical techniques. The neutrino transport is calculated using a multi-energy and multi-species two-moment (M1) scheme with a full complement of neutrino-matter interactions \citep{2006NuPhA.777..356B}. Included are recoil/weak-magnetism corrections to scattering and absorption\citep{2002PhRvD..65d3001H} and axial-vector many body corrections to neutrino-nucleon scattering \citep{2017PhRvC..95b5801H}. Fluid-velocity and general relativistic effects are included in the neutrino transport to lowest order; for the hydrodynamics, a general-relativistic correction is added in the monopole component of the gravitational potential following the method in \citet{2006A&A...445..273M}. We refer readers to \citet{2019ApJS..241....7S} for more details on the characteristics and capabilities of the code.

During the last several years, we have investigated many aspects of CCSN dynamics by performing CCSN simulations in both 2D \citep{2016ApJ...831...81S,2017ApJ...850...43R,2018MNRAS.477.3091V} and 3D \citep{2019MNRAS.482..351V,2019MNRAS.485.3153B,2020MNRAS.491.2715B,2019MNRAS.489.2227V,2019MNRAS.490.4622N,2020MNRAS.492.5764N,2020ApJ...901..108V}. In our new 2D simulations, the same input physics is employed, but we simulate to $\sim 4$ s post-bounce for more than a dozen progenitors. The hydrodynamical aspect of the new results is briefly summarized in \citet{2021Natur.589...29B}. These longer-term simulations enable us to estimate fundamental observables, such as the explosion energy and $^{56}, {\rm Ni}$ mass\footnote{\citet{2020arXiv201005615S} suggest that simulations to at least $\gtrsim 2$ s after core bounce are required to make robust estimates of $^{56} {\rm Ni}$ mass.}, NS mass, and explosive nucleosynthesis. This paper focuses on a detailed analysis of the neutrino signals based on these 2D models.

\begin{figure}
  \rotatebox{0}{
    \begin{minipage}{1.0\hsize}
        \includegraphics[width=\columnwidth]{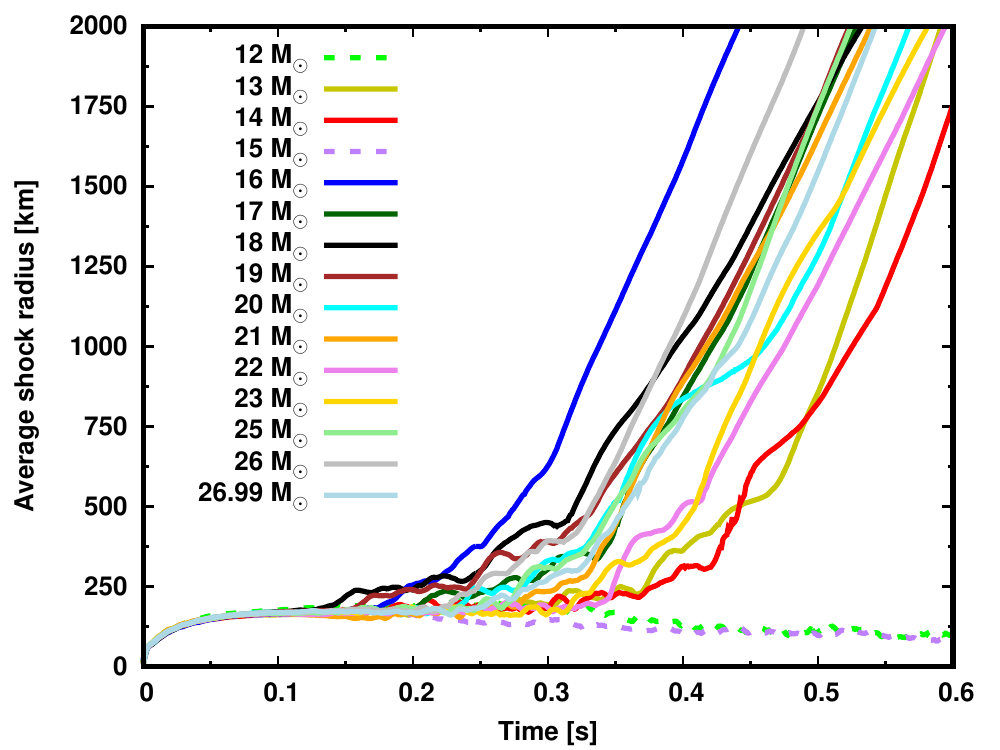}
    \caption{The time trajectory of angle-averaged shock radii for our CCSN models. The solid and dotted lines represent 
models which either succeed or fail.}
    \label{graph_shockevo_2Dlong}
  \end{minipage}}
\end{figure}

We start our CCSN simulations at the onset of gravitational collapse and employ matter profiles at the presupernova phase computed in \citet{2018ApJ...860...93S}. We note that the progenitor models are different from those used in our previous 3D CCSN simulations \citep{2020MNRAS.491.2715B}, except for the 25~$M_{\sun}$ model. In this study, we analyze 15 models over a mass range of $12$ to $26.99 M_{\sun}$. It should be mentioned that the models with low-mass progenitors are not considered here, since the multi-D effects are less prominent for them \citep[see also][]{2021MNRAS.500..696N}. We witness successful explosions in most of our CCSN models, except for the 12 and 15~$M_{\sun}$ progenitors, seen in the time evolution of the angle-averaged shock radii displayed in Fig.~\ref{graph_shockevo_2Dlong}. For all exploding models, shock revival occurs at $<0.5$~s and the models produce observationally expected explosion energies and nucleosynthesis \citep[see][for more details.]{2013ApJ...771...27Y}\footnote{We note, however, that the non-exploding progenitors behave differently \citep[see, e.g.,][]{2019MNRAS.485.3153B}.}. The primary cause of the ``failure" of the non-exploding models seems to be their less prominent Si/O interfaces and their shallower initial density profiles \citep[see Fig. 1 in][]{2021Natur.589...29B}. This trend is consistent with what we found in our previous study \citep{2018MNRAS.477.3091V}.

These self-consistent simulations help us identify the ingredients that characterize the neutrino signals and their dimensional dependence. In the neutrino analysis of our 3D models \citep{2021MNRAS.500..696N}, we concluded that PNS convection is one of the major reasons for the differences of their neutrino signals from those of the 1D models. There is a caveat, however; this conclusion may be valid only in the early post-bounce phase ($\lesssim 1$s) and it depends on the progenitor. As a matter of fact, the vigor of PNS convection strongly depends on epoch and progenitor \citep[see also][]{2020MNRAS.492.5764N}, implying that the impact of PNS convection on the neutrino signals in the late phase is still uncertain. Hence, we address this issue in this paper. We also pay attention to the role of asymmetric mass accretion onto the PNS in the neutrino signals. It should be stressed that the dynamics of mass accretion onto the PNS in multi-D simulations is qualitatively different from that in 1D \citep[see also][]{2010ApJ...725L.106W}. This is also related to how and when the shock wave is revived, indicating that the neutrino emissions bear the stamp of the matter dynamics during the post-shock-revival phases. As we will discuss in Sec.~\ref{sec:neutemiCCSNsource}, the temporal variation of the neutrino signals contains such a hydrodynamical information.

Importantly, the outcome in the non-exploding 12- and 15~$M_{\sun}$ models should not be considered definitive. This is simply because there remain uncertainties in CCSN simulations. One such uncertainty is the stellar evolution model including its rotational, magnetic field, and multi-D stellar profiles. Another is the need for improved treatment of general relativity, multi-angle neutrino transport, and neutrino-matter interactions and of an assessment of the potential role of neutrino oscilllations. Numerical methods and grid and group resolution also affect the final outcome. Therefore, we need to keep in mind such uncertainties as we proceed with the following analysis. Nevertheless, our simulations provide state-of-the-art CCSN models, and non-exploding models still provide distinctive and diagnostic characteristics.

\subsection{Detector simulations}\label{sec:detesimu}
Based on our 2D CCSN models, we estimate event counts in some representative terrestrial neutrino detectors. The method is essentially the same as that used in our neutrino analysis of our 3D CCSN models. We refer readers to \citet{2021MNRAS.500..696N} for the details of our method\footnote{See also \citet{2018MNRAS.480.4710S}, although our analysis pipeline here is slightly different from that used in this paper.}. We now briefly describe the essential elements of our method.

We employ the detector software, SNOwGLoBES\footnote{The software is available at \url{https://webhome.phy.duke.edu/~schol/snowglobes/}.}, to estimate the neutrino counts. In SNOwGLobes, cross section and detector responses in various detector types and reaction channels are provided. Assuming a distance to the CCSN source\footnote{In this paper, we assume that the distance is 10 kpc, unless otherwise stated.} and a neutrino oscillation model (see Sec.~\ref{sec:neutosci}), our analysis starts by constructing mock data of flavor- and energy-dependent neutrino fluxes (fluences) at the Earth by using the neutrino data from our 2D CCSN simulations. We focus only on the angle-averaged neutrino signals, which are very similar as those in our 3D models \citep[see, e.g.,][]{2019MNRAS.482..351V,2019MNRAS.490.4622N}, indicating that they are appropriate stand-ins for 3D. On the other hand, the angular variation in the neutrino signals of our 2D models is not as accurate as in 3D, since the 2D simulations are artificially axisymmetric. Hence, we postpone a detailed analysis of the angular dependence of neutrino signals until longer-term 3D simulations are available. We refer readers to \citet{2021MNRAS.500..696N}, in which the detailed analysis of the angular dependence in the early post-bounce phase ($\lesssim 1$~s) was in fact explored.

Since our CCSN code, F{\sc{ornax}}, is equipped with multi-energy (spectral) and multi-species neutrino transport, the energy spectrum of each flavor of neutrino can be obtained without any artificial spectral prescriptions\footnote{The energy spectrum of neutrinos from CCSN has been frequently been fit by a Gamma distribution with an average energy and a pinching parameter \citep[see, e.g.,][]{2003ApJ...590..971K}, which is very useful for the spectrum analysis of neutrino signals with statistical methods \citep[see, e.g.,][]{2002PhLB..547...37B,2002PhLB..542..239M,2008JCAP...12..006M,2017JCAP...11..036G,2014PhRvD..89f3007L,2016PhRvD..94b3006L,2018JCAP...04..040G,2018PhRvD..97b3019N}. However, such an analytic fit is not particularly accurate; hence, energy spectra extracted from multi-group CCSN simulations are preferable.}. Note that our CCSN simulations do not distinguish mu- and tau- neutrinos (and their anti-particles), which are, hence, collectively treated as ``heavy" leptonic neutrinos ($\nu_x$) in the signal analysis\footnote{Note that we do distinguish the heavy leptonic neutrinos from their anti-partners at the Earth. This is because they undergo different flavor conversions (see also Sec.~\ref{sec:neutosci}).}. It should be noted, however, that their classical neutrino emissions are not identical in reality due to slightly different neutrino-matter interactions. Indeed, the deviation increases with energy. The collective treatment of heavy leptonic neutrinos is, however, a reasonable approximation for $\lesssim 50$ MeV neutrinos. We note that the detection of CCSN neutrinos will be dominated by neutrinos in the energy range of $\lesssim 20$ MeV, indicating that our bundling approach captures all qualitative trends in the neutrino signals. As a final remark, we note that we focus only on the major interaction channel in each detector, which is enough to determine the overall trends in the neutrino signal. It should be mentioned, however, that other channels would be important for the analysis of higher energy neutrinos; we refer readers to \citet{2020arXiv201015136N} for an analyses in the $> 50$ MeV energy range and including subdominant reaction channels.

In this study, we consider four (five including Hypre-K) representative terrestrial detectors: Super-Kamiokande (SK) \citep{2016APh....81...39A} or Hyper-Kamiokande (HK) \citep{2018arXiv180504163H}, the deep underground neutrino experiment (DUNE) \citep{2016arXiv160105471A,2016arXiv160807853A,2020arXiv200806647A}, the Jiangmen Underground Neutrino Observatory (JUNO) \citep{2016JPhG...43c0401A}, and IceCube \citep{2011A&A...535A.109A}. SK and IceCube are currently in operation; the others are coming online and will be available in several years. A reaction channel with inverse beta decay on protons (IBD-p):
\begin{eqnarray}
\bar{\nu}_e  + p \rightarrow e^{+} + n,
\label{ibdchart}
\end{eqnarray}
is a major reaction channel for neutrinos from CCSN in SK (HK), JUNO, and IceCube. DUNE is, on the other hand, sensitive to $\nu_e$ through a charged-current reaction channel with Argon (CCAre):
\begin{eqnarray}
\nu_e  + {^{40}{\rm Ar}} \rightarrow e^{-} + {^{40}{\rm K}^{*}},
\label{ibdchart}
\end{eqnarray}
which is the major channel for that detector. In this paper, we focus on neutrino event counts for the above two reaction channels.

We assume that SK and HK have identical detector configurations, except for the volume, which is set at $32.5$~ktons and $220$~ktons, respectively\footnote{We note that the fiducial volume of the two detectors is usually $22.5$~ktons and $187$~ktons, respectively, since other volumes are used to reduce the background noise. In the burst events such as CCSNe, however, the background may be negligible \citep[see, e.g.,][]{2021PTEP.2021b3E01M}; hence, we assume that the full inner volume can be used in this study.}. The detector volumes of DUNE, JUNO, and IceCube are assumed to be $40$~ktons, $20$~ktons, and $3.5$~Mtons, respectively. In our detector simulations, we take into account smearing effects due to detector response, as provided by SNOwGLoBES. On the other hand, we ignore Poisson noise in this study\footnote{It should be noted, however, that we take into account Poisson noise when discussing the detectability of temporal variations in the neutrino signals. See Sec.~\ref{sec:signalanalysis} for more details.}. Although this should be taken into account when retrieving the energy spectrum of neutrinos, spectral reconstructions are not the main focus of this paper. For an example of spectrum reconstruction, we refer readers to \citet{2021MNRAS.500..319N}, in which the energy spectra of all flavors of neutrino are retrieved by using data in multiple reaction channels and detectors.

\begin{figure*}
  \rotatebox{0}{
    \begin{minipage}{1.0\hsize}
        \includegraphics[width=\columnwidth]{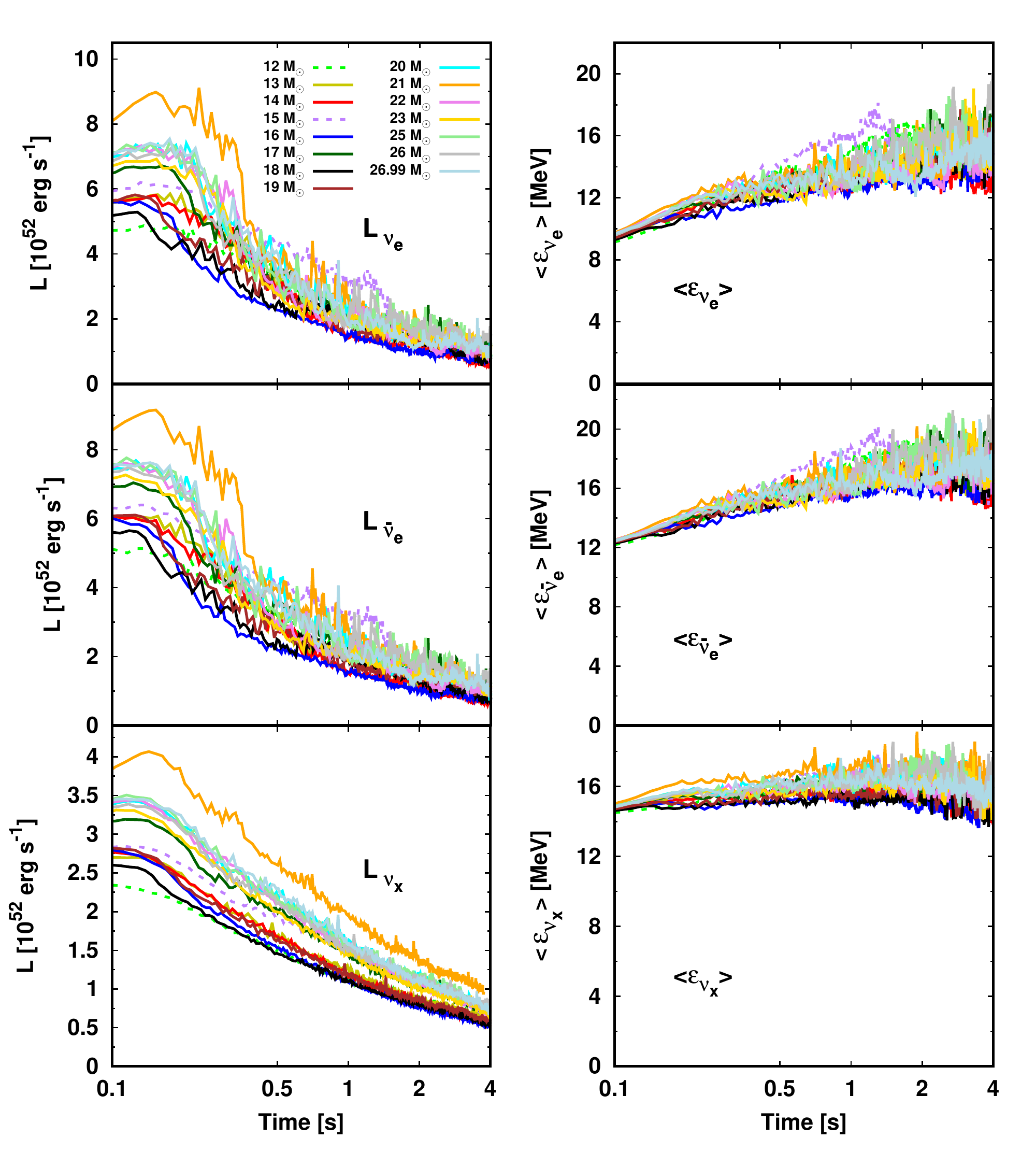}
    \caption{Time evolution of angle-averaged neutrino luminosity (left) and average energy (right). They are evaluated in the laboratory frame and measured at 250 km in the CCSN simulations. From top to bottom, they are $\nu_e$, $\bar{\nu}_e$, and $\nu_x$, respectively. The color represents the progenitor model. The solid and dashed lines distinguish the explosion and non-explosion models, respectively.}
    \label{graph_neutrino_lumi_aveE_2Dlong}
  \end{minipage}}
\end{figure*}

\begin{figure}
  \rotatebox{0}{
    \begin{minipage}{1.0\hsize}
        \includegraphics[width=\columnwidth]{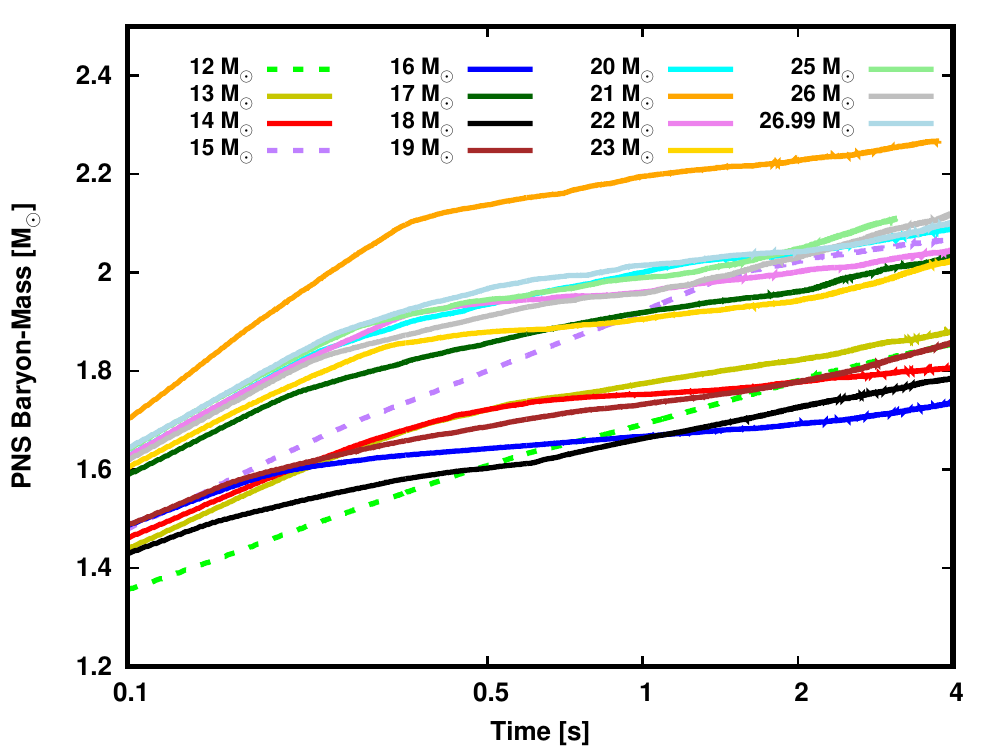}
    \caption{The time evolution of the baryon-mass of the PNS. Similar to Fig.~\ref{graph_shockevo_2Dlong}, the solid and dashed lines distinguish the explosion and non-explosion models.}
    \label{graph_PNSmass_2Dlong}
  \end{minipage}}
\end{figure}

\subsection{Neutrino oscillation model}\label{sec:neutosci}
We employ the simplest, but widely used, neutrino oscillation model. Neutrinos are assumed to execute flavor conversions adiabatically by Mikheyev-Smirnov-Wolfenstein (MSW) effects with matter. The flavor conversion depends on the mass-hierarchy; hence, we study the two cases: normal- and inverted-mass hierarchy. By following \citet{2000PhRvD..62c3007D}, the neutrino flux at the Earth, $F_i$, where the subscript $i$ represents the neutrino flavor, can be computed using those without flavor conversion, $F^{0}_{i}$, as
\begin{eqnarray}
&&F_{e}  = p  F^{0}_{e}  + \left(1- p  \right) F^{0}_{x}, \label{eq:flavconv_nue} \\
&&\bar{F}_{e}  = \bar{p}  \bar{F}^{0}_{e}  + \left(1- \bar{p}  \right) \bar{F}^{0}_{x}, \label{eq:flavconv_nueb} \\
&&F_{x}  = \frac{1}{2}  \left(1- p  \right) F^{0}_{e}
+ \frac{1}{2} \left(1+ p  \right) F^{0}_{x}, \label{eq:flavconv_nux} \\
&&\bar{F}_{x}  = \frac{1}{2}  \left(1- \bar{p}  \right) \bar{F}^{0}_{e}
+ \frac{1}{2}  \left(1+ \bar{p}  \right) \bar{F}^{0}_{x}\, , \label{eq:flavconv_nuxb}
\end{eqnarray}
where $p$ denotes the survival probability which depends upon the neutrino oscillation model and the neutrino mass hierarchy. The upper bar denotes the anti-neutrino quantities. In the case of the normal-mass hierarchy, $p$ and $\bar{p}$ can be written as
\begin{eqnarray}
&& \hspace{-10mm} p = \sin^2 \theta_{13}, \label{eq:p_normal} \\
&& \hspace{-10mm} \bar{p} = \cos^2 \theta_{12} \cos^2 \theta_{13}. \label{eq:barp_normal}
\end{eqnarray}
In the case of the inverted hierarchy, they are
\begin{eqnarray}
&& \hspace{-10mm} p = \sin^2 \theta_{12} \cos^2 \theta_{13}, \label{eq:p_inv} \\
&& \hspace{-10mm} \bar{p} = \sin^2 \theta_{13}\, . \label{eq:barp_inv}
\end{eqnarray}
Following \citet{2017PhRvD..95i6014C}, we adopt the neutrino mixing parameters, $\theta_{12}$ and $\theta_{13}$ as $\sin^2 \theta_{12} = 2.97 \times 10^{-1}$ and $\sin^2 \theta_{13} = 2.15 \times 10^{-2}$, which are the same as those used in \citet{2021MNRAS.500..696N}.

\begin{figure*}
  \rotatebox{0}{
    \begin{minipage}{1.0\hsize}
        \includegraphics[width=\columnwidth]{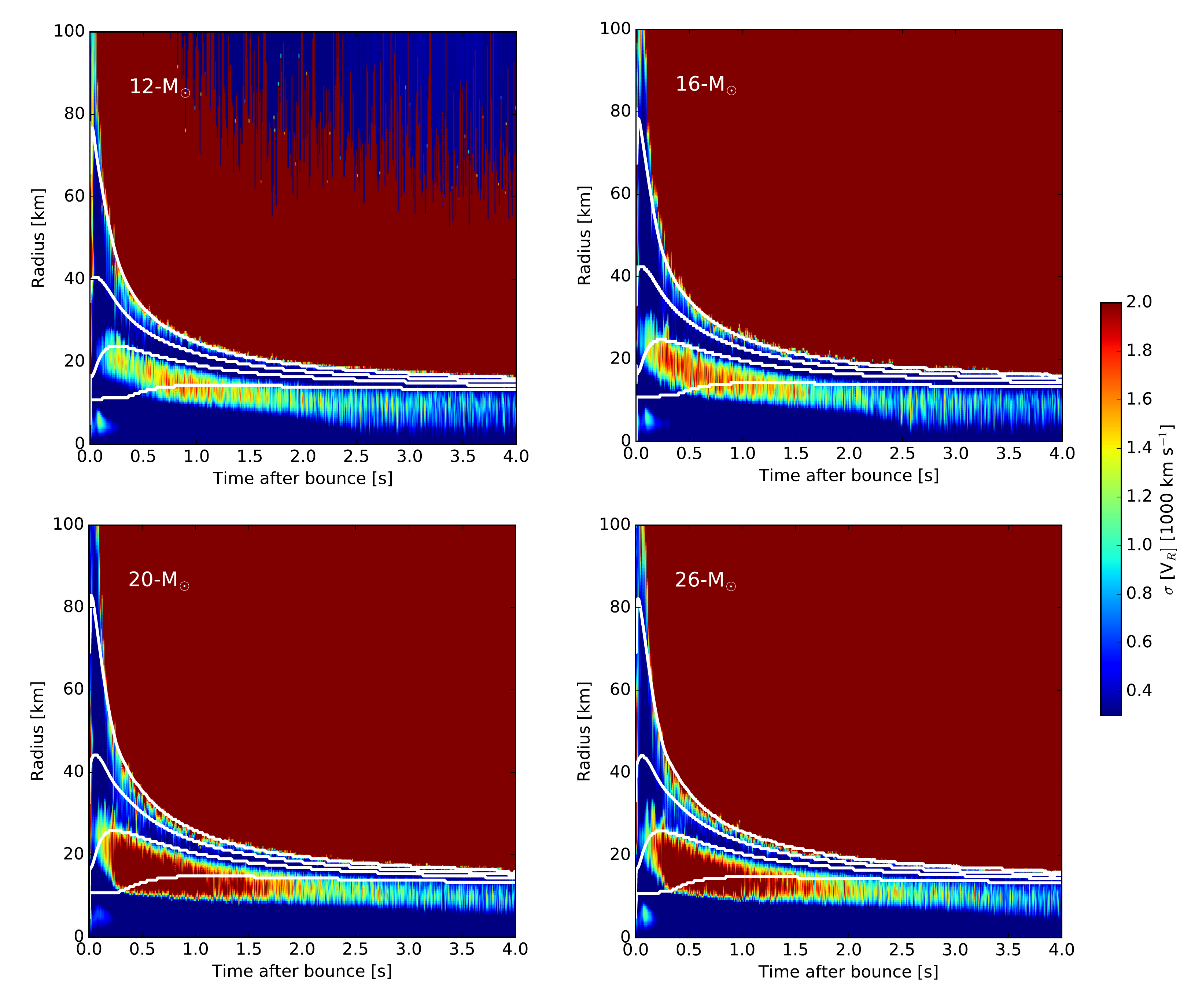}
  \caption{Color map of the angle-averaged lateral speed of fluids displayed as functions of radius and time. We selected four representative models: 12 (top-left), 16 (top-right), 20 (bottom-left) and 26~$M_{\sun}$ (bottom-right). We also display angle-averaged isodensity radii with $10^{11}, 10^{12}, 10^{13}$, and $10^{14} {\rm g/cm}^3$ (from large to small radii) as white lines in each panel.}
    \label{RadiusTimeVt}
  \end{minipage}}
\end{figure*}

\begin{figure*}
  \rotatebox{0}{
    \begin{minipage}{1.0\hsize}
        \includegraphics[width=\columnwidth]{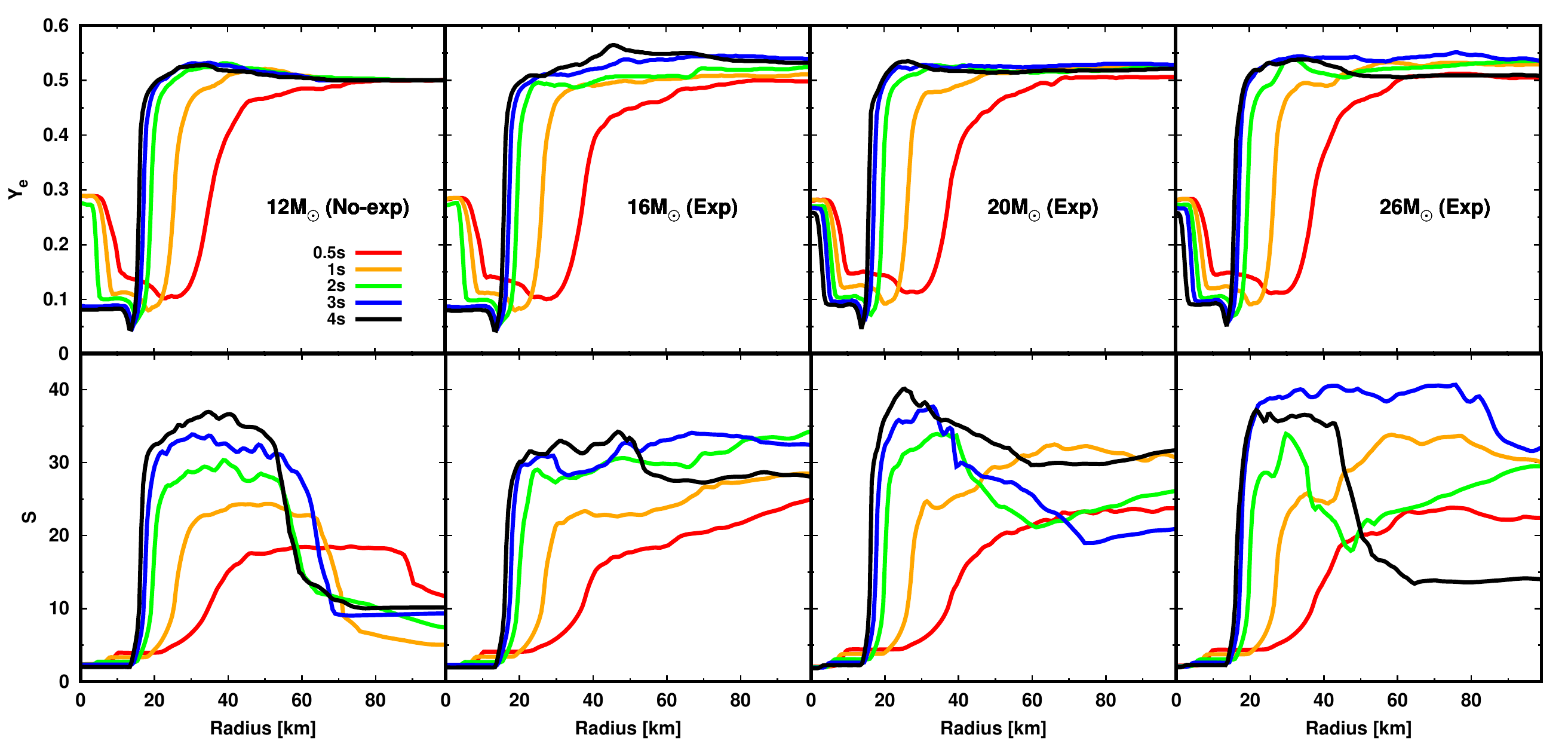}
  \caption{Angle-averaged electron fraction ($Y_e$) and entropy (S) profiles are shown in the top and bottom panels, respectively. From left to right, they are 12, 16, 20, and 26~$M_{\sun}$ models, respectively. Color represents the different time snapshots.}
    \label{graph_Ye_S_angave}
  \end{minipage}}
\end{figure*}

\begin{figure}
  \rotatebox{0}{
    \begin{minipage}{1.0\hsize}
        \includegraphics[width=\columnwidth]{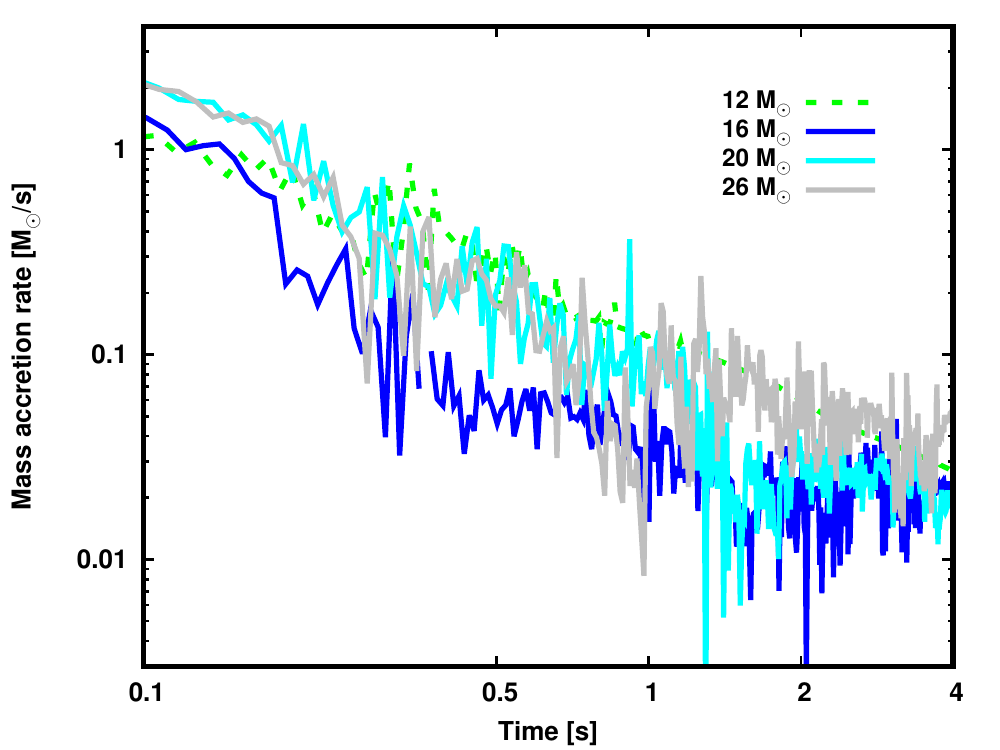}
    \caption{The time evolution of the mass accretion rate measured at $100$~km for selected models, 12, 16, 20, and 26~$M_{\sun}$ models, which are distinguished by color. The line type denotes  exploding (solid) or non-exploding (dashed) models.}
    \label{graph_Massaccretion_2Dlong}
  \end{minipage}}
\end{figure}

\begin{figure}
  \rotatebox{0}{
    \begin{minipage}{0.9\hsize}
        \includegraphics[width=\columnwidth]{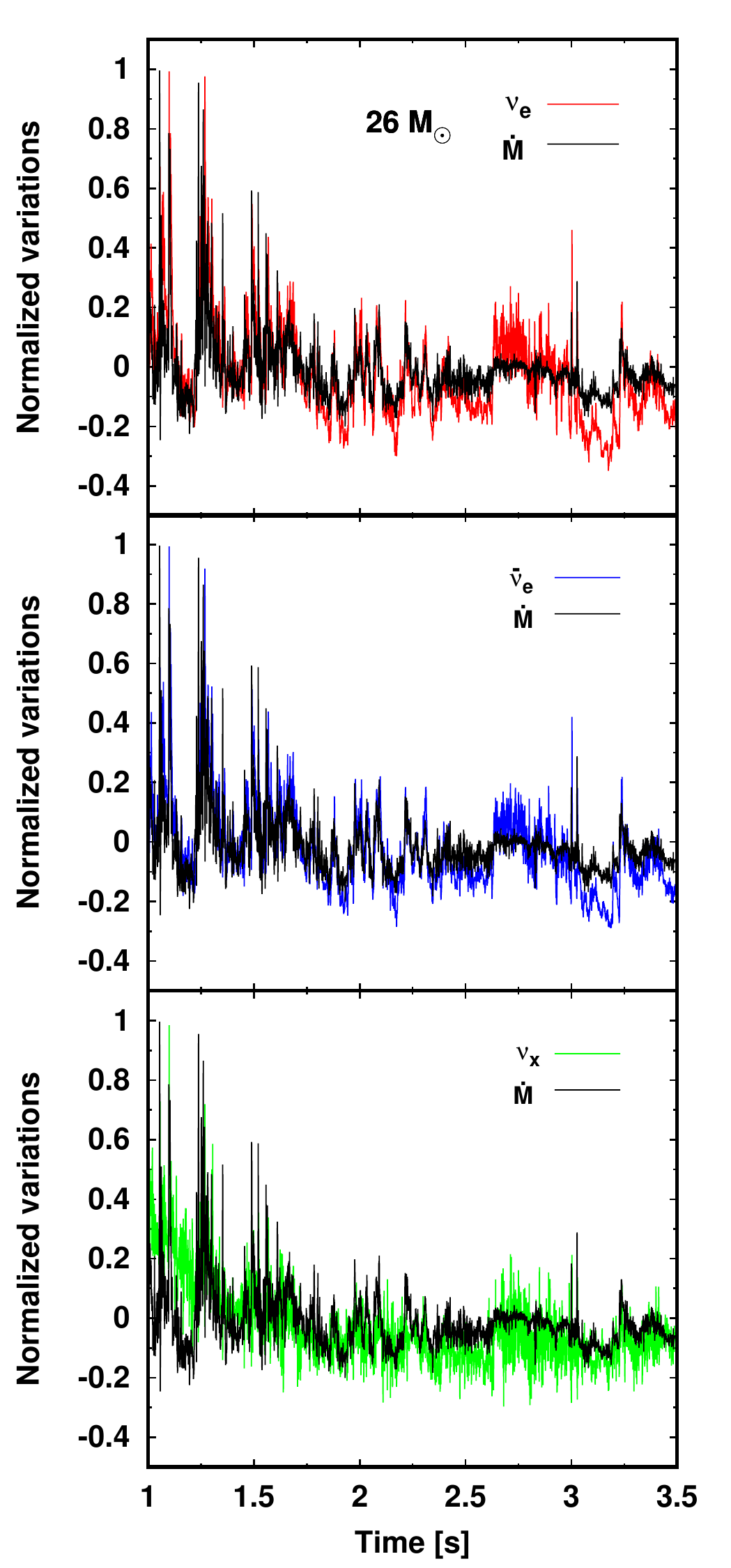}
    \caption{Time variable component of the neutrino luminosity and mass accretion rate as a function of time. From top to bottom, we compare the luminosities of the $\nu_e$, $\bar{\nu}_e$, and $\nu_x$ neutrinos to the mass accretion rate, respectively. The vertical axis is normalized such that the peak amplitude is unity. See Eq.~\ref{eq:defvariable} and text for the definition of the variable.}
    \label{graph_temporalvali_neutrino_mdot}
  \end{minipage}}
\end{figure}

\begin{figure*}
  \rotatebox{0}{
    \begin{minipage}{1.0\hsize}
        \includegraphics[width=\columnwidth]{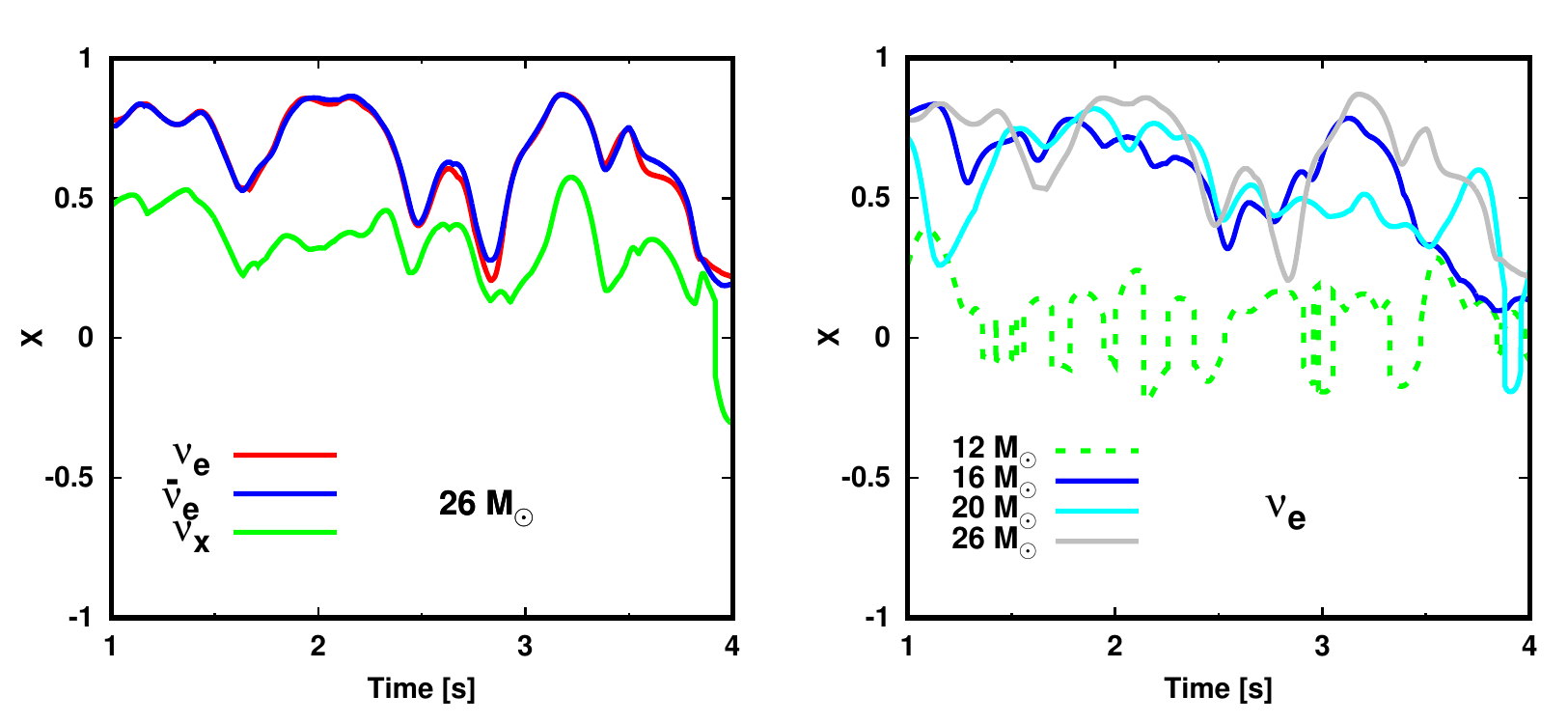}
    \caption{Correlation function of the temporal variation of the neutrino luminosity and mass accretion rate. The left panel shows the result for the 26~$M_{\sun}$ model. The color distinguishes the neutrino species. The right panel displays the same as the left one, but shows the progenitor dependence. In the panel, we display only the result of the $\nu_e$ and $\dot{M}$ correlation. See the text for more details.}
    \label{graph_TimeCorrelation_NeutrinosMdot_shortTerm}
  \end{minipage}}
\end{figure*}

\begin{figure*}
  \rotatebox{0}{
    \begin{minipage}{1.0\hsize}
        \includegraphics[width=\columnwidth]{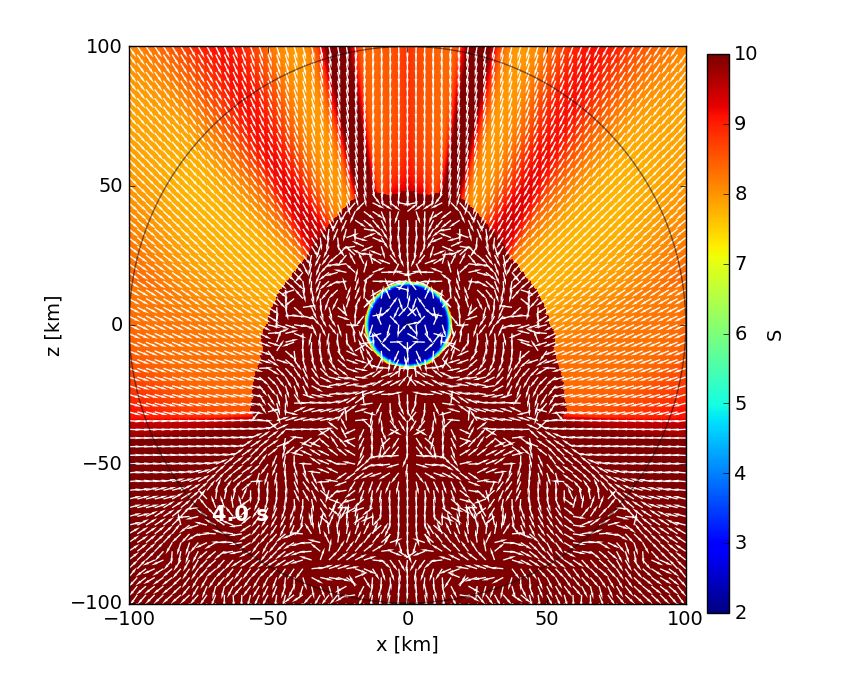}
    \caption{The entropy per baryon in color with fluid velocities in vectors for the 26~$M_{\sun}$ model at $4$~s.}
    \label{graph_secoshockmap}
  \end{minipage}}
\end{figure*}

\section{Results}\label{sec:results}

\subsection{Neutrino emissions at CCSNe}\label{sec:neutemiCCSNsource} 
First, we analyze neutrino emissions in the supernovae. Fig.~\ref{graph_neutrino_lumi_aveE_2Dlong} displays the time evolution of the neutrino luminosities and average energies. Rich flavor-, time-, and progenitor-dependent features emerge. The difference in the  mass accretion rates onto the PNS is primarily responsible for generating this diversity. As evidence, both mass accretion rate and neutrino luminosity (regardless of flavor) during the early post-bounce phase ($\lesssim 0.4$~s) in the 21~$M_{\sun}$ model are the highest among all the models shown. We also note that the higher mass accretion rate builds a more massive PNS (see Fig.~\ref{graph_PNSmass_2Dlong}), which boosts the neutrino luminosity in the late phase. The high mass accretion rate is correlated with the core compactness parameter;
those qualitative trends are consistent with what has been reported in previous studies based on 1D models \citep[see, e.g.,][]{2013ApJS..205....2N,2020ApJ...898..139W,2021arXiv210110624S}. It should also be pointed out that the multi-D effect on the neutrino signal is mild in the very early post bounce phase ($\lesssim 50$~ms) \citep[see also][]{2021MNRAS.500..696N}. This suggests that the neutrino signal in the early phase can be used to constrain the distance to the CCSN progenitor \citep{2021arXiv210110624S}. However, we find
that strong temporal variations emerge in the late phase, and these have not been reported in previous studies. The variation amplitude depends upon progenitor and neutrino flavor. These temporal characteristics are the missing from previous studies. Below, we perform an in-depth analysis of its physical origin.

We first consider the effects of PNS convection on the neutrino signals, since this multi-D fluid instability is a major missing element in 1D models and actually affects the neutrino emissions at $\lesssim 1$ s \citep[see][for more details]{2021MNRAS.500..696N}. We start by analyzing properties of PNS convection itself; Fig.~\ref{RadiusTimeVt} portrays the time evolution of the radial profiles of the angle-averaged tangential fluid speed for some selected models: 12, 16, 20, and 26~$M_{\sun}$. We confirm that PNS convection commonly occurs for all progenitors in the region $10 \lesssim r \lesssim 25$~km at $\lesssim 1$~s, which is the same as seen in 3D models \citep{2020MNRAS.492.5764N}. On the other hand, during the late phase ($\gtrsim 1$ s) PNS convection gradually subsides.

As discussed in \citet{2020MNRAS.492.5764N}, PNS convection is mainly driven by the negative lepton number gradient at the edge of the compact core. Thermal energy and lepton number in the PNS are progressively diminished via neutrino emission. This drives a quasi-steady change in the PNS state towards neutrinoless beta equilibrium at zero temperature. The time-dependent feature of deleptonization and neutrino cooling can be seen in Fig.~\ref{graph_Ye_S_angave}; both electron fraction ($Y_e$) and entropy per baryon ($S$) around the surface of PNS $\sim 10$ km decrease with time. It should be mentioned that PNS convection facilitates its deleptonization and cooling \citep[see also][]{2012PhRvL.108f1103R}, indicating that the quasi-steady evolution differs from that in 1D models. As shown in the top panels of Fig.~\ref{graph_Ye_S_angave}, the location of sharp negative $Y_e$ gradient gradually sinks into the inner region. Eventually, the inner edge of PNS convection reaches the coordinate center (mass center of the PNS), which can be seen in Fig.~\ref{graph_Ye_S_angave} for the 12 and 16~$M_{\sun}$ models in the $3$ and $4$~s snapshots\footnote{The arrival time of the inner edge of PNS convection at the center depends upon the model; the decay of PNS convection tends to take longer for heavier proto-neutron stars.}. Once it reaches the center, the negative lepton number gradient starts to disappear; this corresponds to the time PNS convection nearly ceases. Hence, the vigor of PNS convection becomes weaker with time in the late post-bounce phase.

Based on these results, let us consider the impact of PNS convection on the neutrino emissions. First, we point out that the temporal variations in neutrino emissions are seen even after PNS convection has subsided (at $\gtrsim 2$ s). In addition to this, if PNS convection played the primary role in the temporal variation of the neutrino emissions, all flavors of neutrino would have similar temporal variation. However, we find that the temporal variation of the $\nu_x$ emissions is remarkably weak when compared to that of the other species, regardless of CCSN model (see left panels in Fig.~\ref{graph_neutrino_lumi_aveE_2Dlong}), which is inconsistent with the above argument. For these reasons, we conclude that the late-time temporal variation in neutrino emission is not primarily driven by PNS convection\footnote{We note that PNS convection potentially affects the weak temporal variation in the $\nu_x$s. See below for more details.}.

It is mass accretion onto the PNS that is the cause of the late-time temporal variation in the neutrino signals. This conclusion is buttressed by the fact that we observe long-lasting mass accretion onto the PNS in all models. The mass of the PNS monotonically increases up to the end of all of our simulations, regardless of progenitor (see Fig.~\ref{graph_PNSmass_2Dlong})\footnote{It should be noted that very light progenitors such as the 9~$M_{\sun}$ model, which is not included in our study, may be exceptions, for which mass accretion almost ceases after shock revival. This is mainly due to the steep density gradient outside its progenitor core.}. One may think that this is due simply to weak explosions accompanied by a large amount of fallback accretion. However, our models include cases with strong explosions\footnote{For instance, the explosion energy of the 26~$M_{\sun}$ model is $\sim 2.3 \times 10^{51}$~erg. See \citet{2021Natur.589...29B} for more details.}. This trend is qualitatively different from that observed in 1D models. In 1D, strong explosions unbind most of the post-shock matter above the PNS. As a result, the late-time accretion rate is subtle\footnote{We note that strong fallback accretion may occur even in 1D due to the reverse shock generated by the deceleration of the shock wave in the hydrogen envelope \citep[see, e.g.,][]{1989ApJ...346..847C,2015A&A...577A..48W}. However, this would occur at a very late phase ($\sim$ hours), which is not a phase we consider in this paper.}. On the other hand, shock revival in multi-D occurs rather asymmetrically, indicating that the vigor of shock expansion depends upon the geometry. It is, hence, possible to have weak shock expansion in some directions even when the overall explosion is strong. This actually happens in our CCSN models; for instance, the 26~$M_{\sun}$ model has a strong dipolar explosion, and it is also accompanied by large amounts of early fallback accretion around the equator. We note that this trend is common in multi-D CCSN models, and the mass inflow can last for more than a few seconds \citep[see, e.g.,][]{2006ApJ...640..891Y,2009ApJ...699..409F,2010ApJ...725L.106W,2018ApJ...852L..19C,2019MNRAS.484.3307M,2020MNRAS.495.3751C}\footnote{It should be noted, however, that the detailed accretion structures would depend on dimension. In 3D, the shock morphology is generally more spherical than in 2D, indicating that the asymmetry of the accretion flows may be reduced.}. Since asymmetric shock expansion and fluid-instabilities alter matter motion in the post-shock region, the accretion inflow onto the PNS is highly disorganized, causing the temporal variation of the neutrino emissions we witness.

Motivated by the above considerations, we take a look at the time evolution of the accretion rate onto the PNS, which is displayed in Fig.~\ref{graph_Massaccretion_2Dlong} for some selected models. As expected, this plot clearly displays both the long-lasting accretion onto PNS and the strong temporal variation for the explosion models (solid lines). The amplitudes of temporal variation are roughly tens of percent of the short-time-average (quasi-steady) component (see below for the definition of the quasi-steady component). On the other hand, the temporal variation is rather mild in the 12~$M_{\sun}$ non-exploding model (see below for more details). At first glance, the amplitude of the temporal variations is positively correlated with the neutrino luminosities. For instance, the mass accretion rate of the 26~$M_{\sun}$ model evinces large temporal variations at $\sim 1$~s and the neutrino luminosities in the same model fluctuate with time strongly in phase.

Let us now explore the correlation. First, we first compute the quasi-steady component of the neutrino luminosity and mass accretion rate to extract their (temporal) variation component. We define the quasi-steady component ($A^{\rm qs}$) with respect to an arbitrary time-dependent quantity $A$ as,
\begin{eqnarray}
A^{\rm qs} (t) = \frac{1}{\Delta t} \int^{t+0.5 \Delta t}_{t-0.5 \Delta t} d \tau \hspace{0.5mm} A(\tau) , \label{eq:defquasisteady}
\end{eqnarray}
and we set $\Delta t$ to $300$ ms in this study\footnote{We checked the dependence of the choice of $\Delta t$, and we confirmed that our results are insensitive to this choice, unless $\Delta t$ is $\lesssim 100$ ms.}. By using $A_{\rm qs}$, we define the time variable component ($A^{\rm tv}$):
\begin{eqnarray}
A^{\rm tv} (t) = A(t) - A^{\rm qs} (t). \label{eq:defvariable} 
\end{eqnarray}
Fig.~\ref{graph_temporalvali_neutrino_mdot} shows $A^{\rm tv}$ as a function of time for neutrino luminosities and mass accretion rates in the 26~$M_{\sun}$ model. Note that the vertical axis is normalized so that the peak amplitude of $A^{\rm tv}$ of each quantity in the interval $1-4$ s is set to unity. This plot clearly shows a strong positive correlation of the temporal variation between $\nu_e$ ($\bar{\nu}_e$) luminosity and mass accretion rate. We can also see that the temporal variations of $\nu_x$ are correlated with the mass accretion rate, albeit more weakly than $\nu_e$ and $\bar{\nu}_e$. 

\begin{figure*}
  \rotatebox{0}{
    \begin{minipage}{1.0\hsize}
        \includegraphics[width=\columnwidth]{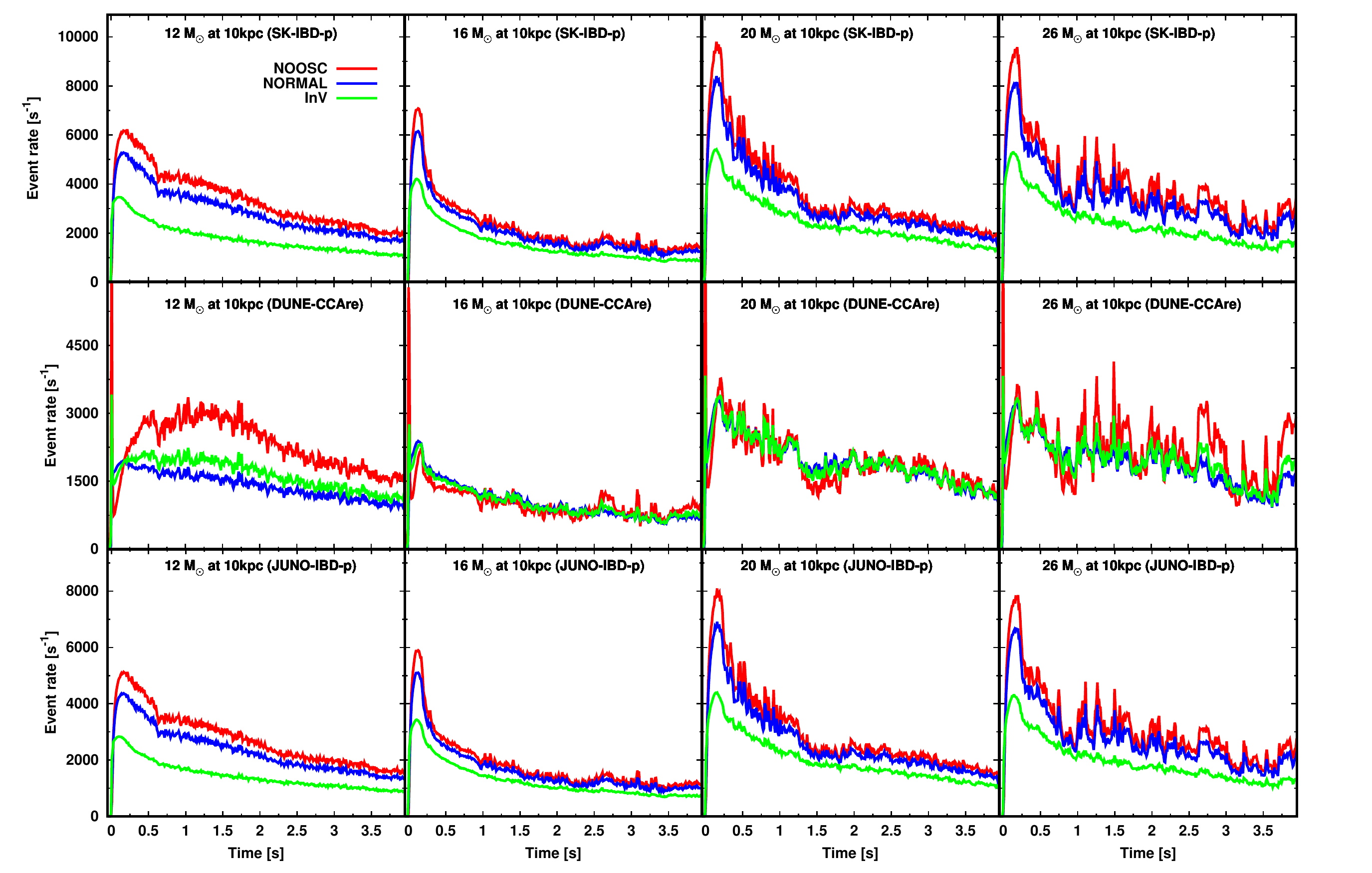}
    \caption{Event rates detected in the major reaction channels for each detector as a function of time. The results for SK, DUNE, and JUNO are displayed from top to bottom. From left to right, we show different CCSN models: 12, 16, 20, and 26~$M_{\sun}$. The color distinguishes the neutrino oscillation models: red (no oscillations), blue (normal mass hierarchy with adiabatic MSW), green (inverted mass hierarchy with adiabatic MSW).}
    \label{graph_evoEverate_2Dlong}
  \end{minipage}}
\end{figure*}

\begin{figure*}
  \rotatebox{0}{
    \begin{minipage}{1.0\hsize}
        \includegraphics[width=\columnwidth]{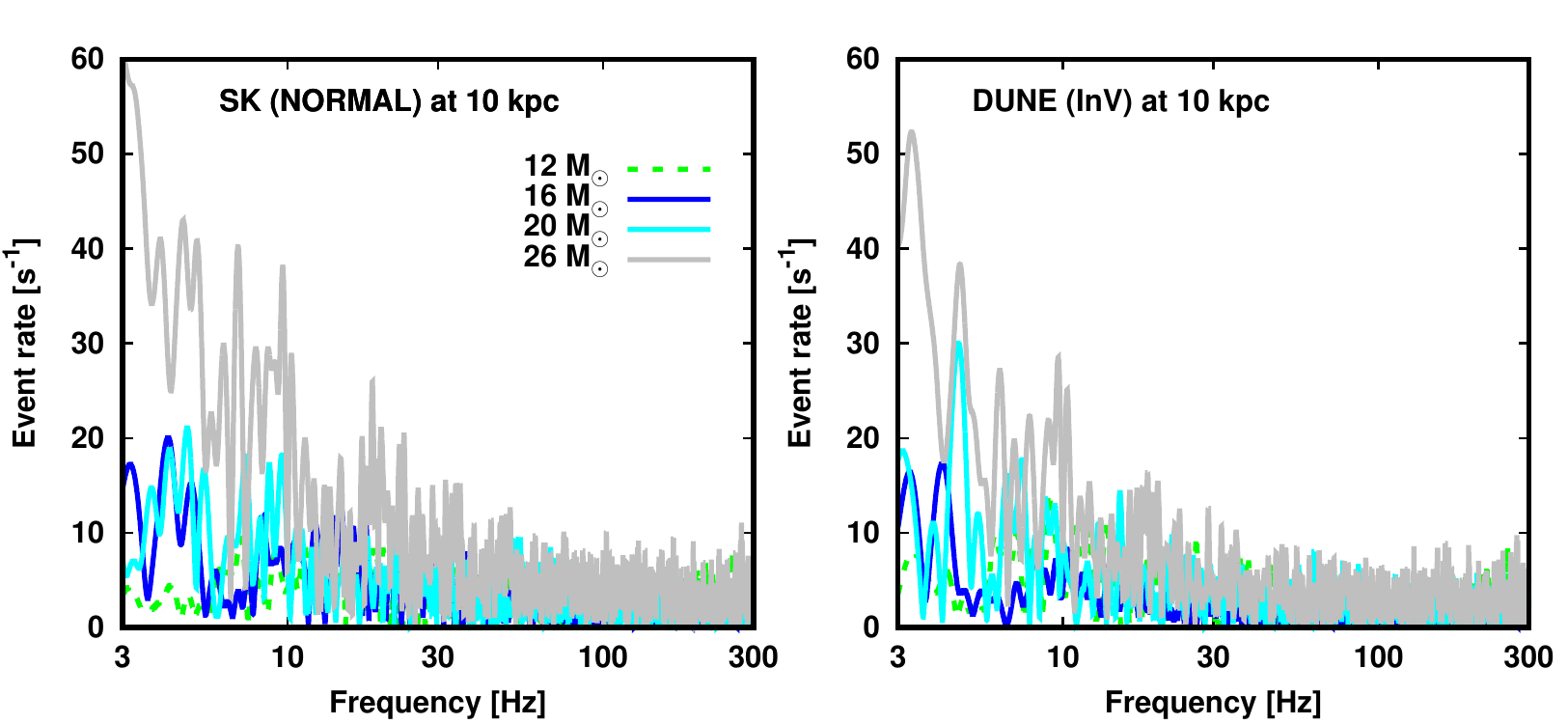}
    \caption{The Fourier transform of the event rate in the late post-bounce phase. The time window is chosen from $1$~s after the core bounce to the end of a simulation. In the left (right) panel, we show the case with SK (DUNE) in normal (inverted) mass hierarchy for CCSNe at $10$~kpc. The color distinguishes CCSN models. The dashed line indicates the 12~$M_{\sun}$ non-exploding model, while solid lines are used for exploding models.}
    \label{graph_TimeFourier}
  \end{minipage}}
\end{figure*}

To assess the correlation more quantitatively, we compute a normalized correlation function of the temporal variation between each species of neutrino and the mass accretion rate \citep[see][for the similar definition of the correlation function]{2017ApJ...851...62K}, which can be written as:
\begin{eqnarray}
{\rm X} (t,\Delta {\rm T}) = \frac{ {\rm Y}_{\nu \dot{M}} (t,\Delta {\rm T})  }{ {\rm Y}_{\nu} (t,\Delta {\rm T})   \times {\rm Y}_{\dot{M}} (t) }, \label{eq:Cordef}
\end{eqnarray}
where
\begin{eqnarray}
&&{\rm Y}_{\nu \dot{M}} (t,\Delta {\rm T}) = \int  d \tau \hspace{0.5mm}  H(t - \tau) A^{\rm tv}_{\nu}( \tau + \Delta {\rm T} ) A^{\rm tv}_{\dot{M}} (\tau), \nonumber \\
&&{\rm Y}_{\nu} (t,\Delta {\rm T}) = \sqrt{ \int  d \tau \hspace{0.5mm}  H(t - \tau) \left(A^{\rm tv}_{\nu}( \tau + \Delta {\rm T} ) \right)^2 }, \nonumber \\
&&{\rm Y}_{\dot{M}} (t) = \sqrt{ \int  d \tau \hspace{0.5mm}  H(t - \tau) \left(A^{\rm tv}_{\dot{M}}( \tau  ) \right)^2 }. \label{eq:CordCompodef}
\end{eqnarray}
In the expression, $H$ denotes the Hann window function. The size of the time window is set to $300$ ms. $\Delta {\rm T}$ represents the time delay of the response of the neutrino luminosity to the temporal variation of the mass accretion rate. Since $\Delta {\rm T}$ is not known a priori, it is varied in the range of $0-10$ ms in this study\footnote{We note that $\Delta {\rm T}$ also depends on where we measure neutrino signals ($250$ km in this study) and mass accretion rate (here at $100$ km). Taking into account neutrino propagation, the actual delay time of the response would be smaller than $\Delta {\rm T}$.}. We note that our employed correlation function may be improved by more sophisticated prescriptions to extract the quasi-steady component \citep[see, e.g.,][]{2018NJPh...20g3005C}. However, our method suffices for the purposes of this paper; indeed it captures the qualitative trend of the correlation, as we now show.

Fig.~\ref{graph_TimeCorrelation_NeutrinosMdot_shortTerm} portrays the time evolution of the correlation function of the temporal variations between neutrino luminosity and mass accretion rate. In the plot, $\Delta {\rm T}$ is chosen so that the absolute value of the correlation function is a maximum. Roughly speaking, $\Delta {\rm T}$ is $\sim 2$ ms for most of the models and all models shown have strong correlations. We focus on the 26~$M_{\sun}$ model in the left panel and show the flavor-dependent feature of the correlation function. We confirm a strong correlation of the temporal variation between the $\nu_e$ ($\bar{\nu}_e$) luminosity and mass accretion rate. On the other hand, $\nu_x$ has the least correlation among them. The weak correlation may indicate that PNS convection affects the temporal variations, albeit subdominantly. In the right panel of Fig.~\ref{graph_TimeCorrelation_NeutrinosMdot_shortTerm}, we show the progenitor dependence for selected models, focusing on $\nu_e$. The positive correlation is clearly shown in other CCSN models, except for the 12~$M_{\sun}$ model. Hence, we conclude that temporal variation in the  mass accretion rate onto the PNS is the most influential cause of the temporal variations in the neutrino signals.

There are a few caveats to this conclusion. Although we reveal that inhomogeneous mass accretion flows play a dominant role in determining the temporal characteristics of neutrino signals, the correlation weakens with time; indeed, it is less than 0.5 for all models at $4$ s (see Fig.~\ref{graph_TimeCorrelation_NeutrinosMdot_shortTerm}). Meanwhile, temporal variations still exist at that time (see Fig.~\ref{graph_temporalvali_neutrino_mdot}). The weak correlation with the mass accretion rate indicates that there is another driver creating temporal variation in neutrino signals. We suggest this is due to a fluid instability right above the PNS\footnote{It should be mentioned that the fluid instability is different from the PNS convection, which happens deeper inside the PNS.}. Fig.~\ref{graph_secoshockmap} displays the entropy distribution in the central region for the 26~$M_{\sun}$ model at $4$~s. As shown in the figure, the accretion shock wave emerges in the northern hemisphere. The shock wave fluctuates on a timescale of a few ms, and the overall behavior shows change on a $\sim 100$~ms timescale. On the other hand, shock dynamics (instability) strongly depends upon dimension, indicating that it is unclear if our finding is generic in 3D. Hence, we postpone the detailed study of this dynamics to future work. Nevertheless, it may be that shock instability by fallback accretion may emerge much earlier than previously thought \citep[see][]{1989ApJ...346..847C,1992ApJ...395..592H}.

We also caution that the small temporal variation found in non-explosing models may in part be due to numerical artifacts introduced by using 2D. This is because the spiral Standing Accretion Shock Instability (SASI) is, in general, observed in non-exploding 3D models, which induces a strong quasi-periodic temporal variation in the neutrino signals \citep[see also][]{2021MNRAS.500..696N}. We note that 2D models are not capable of capturing the non-axisymmetric mode, indicating that the spiral SASI is suppressed artificially. To the contrary, we speculate that the temporal variation found in our 2D explosion models may be overestimated compared to that in 3D. This is attributed to the fact that the explosion geometry may be too asymmetric in 2D, which overestimates the asymmetry in mass accretion rates and neutrino signals. Addressing these issues also requires sophisticated 3D long-term simulations, which are postponed to future work. Having in mind these caveats, we move on to the analysis of neutrino signals at the Earth.

\subsection{Neutrino signals at the Earth}\label{sec:signalanalysis}

Fig.~\ref{graph_evoEverate_2Dlong} displays the angle-averaged neutrino event count as a function of time for selected models: 12, 16, 20, and 26~$M_{\sun}$. In the early phase ($\lesssim 1$ s), we see the same trends seen in 3D models \citep{2021MNRAS.500..696N}; the difference in the accretion component of the neutrino luminosity accounts for most of the the progenitor dependence of the event count rate, regardless of detector. The dependence on neutrino oscillation model in 2D models is exactly the same as in 3D. At the late phase, some new features appear in the neutrino signals. First, the quasi-steady component of the neutrino event count rate becomes less sensitive to neutrino oscillation model, which is consistent with that reported in previous 1D studies \citep[see, e.g.,][]{2019ApJ...881..139S}\footnote{Although the trend is common, the event counts in 1D models are different from those in multi-D models \citep[][]{2021MNRAS.500..696N}. In the early phase, PNS convection is mainly responsible for this difference. In the late phase, the difference is remarkable, in particular for successful explosion models, in which long-lasting asymmetric mass accretion in multi-D models boosts the event counts (see also Sec.~\ref{sec:neutemiCCSNsource}).}. This is attributed to the fact that neutrino emission at the source evolves into a common luminosity and spectrum across the three flavors. This can be understood as follows. The neutrino emissions in the late phase have quasi-thermal (Fermi-Dirac) spectra and are characterized roughly by a temperature and chemical potential at the neutrinosphere. The neutrinosphere is less sensitive to flavor due to the sharp density gradient in the PNS envelope\footnote{The PNS envelope contracts with time due to energy loss by neutrino emissions. The sharp density gradient at the outer PNS boundary in the late phase can be seen in Fig.~\ref{RadiusTimeVt}. As shown in this figure, the isodensity radii for different densities converge to the same radius, indicating that the density gradient is very steep.}, indicating that the difference in neutrino temperature among flavors is small. Furthermore, lepton loss from the PNS by neutrino emission reduces $Y_e$ inside of the PNS, which makes the chemical potential of $\nu_e$ (and $\bar{\nu}_e$) neutrinos approach zero. Since the chemical potential of heavy leptonic neutrinos is zero (unless on-shell muons appear in matter \citep{2017PhRvL.119x2702B,2020PhRvD.102l3001F}), the difference in chemical potential among the three flavors of neutrino reduces with time. For these reasons, all the neutrinos evolve towards the identical spectrum.

There is significant diversity in the count rates as a function of both neutrino oscillation and CCSN models. As shown in Fig.~\ref{graph_evoEverate_2Dlong}, the event counts at $\gtrsim 1$~s strongly vary with time in SK, HK, and JUNO for the 20 and 26~$M_{\sun}$ models, unless the neutrino mass hierarchy is inverted. This is consistent with our discussion in Sec.~\ref{sec:neutemiCCSNsource} that $\bar{\nu}_e$ emissions at the CCSN source strongly vary in time for these models (see Sec.~\ref{sec:neutemiCCSNsource})\footnote{On the other hand, for the inverted mass hierarchy the SK, HK, and JUNO signals are sensitive to the $\nu_x$ emissions at the supernova. Therefore, the event count time variability is in this case least between the different oscillation models.}. Similarly, the event count rate in DUNE is strongly time-variable, unless the neutrino mass hierarchy is normal. This behavior originates from the strong time variations in the $\nu_e$ emissions at the supernova source. The time variability is less remarkable for the 12 and 16~$M_{\sun}$ models, regardless of neutrino oscillation model. As mentioned in Sec.~\ref{sec:neutemiCCSNsource}, the non-exploding models tend to have weak time variability (ignoring a possible 3D spiral SASI), and this is responsible for the weak variations we see in the 12~$M_{\sun}$ model. For the 16~$M_{\sun}$ model, the temporal variation in the  mass accretion rate is strong (see Fig.~\ref{graph_Massaccretion_2Dlong}). However, this model has the smallest mean mass accretion rate among our models (see Figs.~\ref{graph_PNSmass_2Dlong}~and~\ref{graph_Massaccretion_2Dlong}). This implies that the accretion component of the neutrino luminosity is also small, indicating that the temporal variation is smeared out by the core diffusion component of the neutrino luminosity. We note, however, that the positive correlation of the temporal correlation between neutrino signals and mass accretion rate remains strong even in those models with weak time variability in the neutrino signals (see Fig.~\ref{graph_TimeCorrelation_NeutrinosMdot_shortTerm}).

To see the temporal structure more clearly, Fig.~\ref{graph_TimeFourier} portrays the Fourier transform of the event rate after $1$~s in the case of the SK detector with the normal mass hierarchy (left) and the DUNE detector with the inverted mass hierarchy (right). Although there emerge no strongly characteristic time frequencies, we find that the low frequencies ($\lesssim 20$~Hz) dominate the temporal structure. The dominance by low-frequency variations supports the conclusion that temporal variations in the neutrino signals are not primarily driven by matter dynamics in the vicinity of the PNS, but rather by external factors such as accretion flows onto the PNS.

Below, we assess the detectability of these temporal variations. We note that in reality the time variations of the event counts may be smeared out by various sources of noise; hence, we take them into account in this discussion. The goal of this estimation is to determine the minimum time bin ($\Delta {\rm T}_{\rm bin}$) for which the time variations dominate the noise, thus providing the highest resolution possible for the time and frequency when performing the Fourier analysis. Based on this estimation, we discuss the detectability of temporal variations.

We start by extracting the quasi-steady component of the event count, which can be done by using Eq.~\ref{eq:defvariable}, i.e.,
\begin{eqnarray}
n^{\rm qs} (t) = \frac{1}{\Delta t} \int^{t+0.5 \Delta t}_{t-0.5 \Delta t} d \tau \hspace{0.5mm} n(\tau)\, , \label{eq:defquasisteady}
\end{eqnarray}
where $n$ and $n^{\rm qs}$ denote the (raw) event count and its quasi-steady component, respectively. The selection of $\Delta t$ can be rather arbitrary, but we suggest that a few hundreds milliseconds is appropriate for extracting the quasi-steady component (see Sec.~\ref{sec:neutemiCCSNsource}).

We first consider the cases of SK (HK), DUNE, and JUNO, in which Poisson noise is the dominant source of detector noise. The Poisson noise in event counts with a time window of $\Delta {\rm T}_{\rm bin}$ can be estimated as
\begin{eqnarray}
N^{\rm noise} (t) \sim \left(  n^{\rm qs} (t) \hspace{0.5mm} \Delta {\rm T}_{\rm bin}  \right)^{0.5}.
\label{eq:noise}
\end{eqnarray}
The temporal component of the neutrino signal can be estimated as
\begin{eqnarray}
N^{\rm tv} (t) \sim | n(t) - n^{\rm qs} (t) | \Delta {\rm T}_{\rm bin} \equiv \alpha (t) \hspace{0.5mm} n^{\rm qs} (t) \hspace{0.5mm} \Delta {\rm T}_{\rm bin}\, ,
\label{eq:tvsignal}
\end{eqnarray}
where $\alpha$ denotes the degree of temporal variation. Thus, the signal-to-noise ratio (SNR) can be given as
\begin{eqnarray}
{\rm SNR} (t) = \frac{N^{\rm tv} (t) }{N^{\rm noise} (t) } \sim \alpha (t) \left( n^{\rm qs} (t) \hspace{0.5mm} \Delta {\rm T}_{\rm bin} \right)^{0.5}.
\label{eq:SNRatio}
\end{eqnarray}
We estimate the required $\Delta {\rm T}_{\rm bin}$ by inserting a typical value for each parameter. $n^{\rm qs}$ is higher than $\gtrsim 2000$ for SK, DUNE, and JUNO (see Fig.~\ref{graph_evoEverate_2Dlong}) (the distance of CCSN is assumed to be $10$ kpc). $\alpha$ depends on neutrino oscillation and progenitor models, but it is $\sim 0.2$ in such optimistic cases as the 26~$M_{\sun}$ model. If we set the threshold SNR to 5, $\Delta {\rm T}_{\rm bin}$ can be estimated to be
\begin{eqnarray}
\Delta {\rm T}_{\rm bin} \sim 300 \hspace{0.5mm} [{\rm ms}] \left( \frac{ {\rm SNR} }{ 5 } \right)^2 \left( \frac{ \alpha }{ 0.2 } \right)^{-2}    \left( \frac{ n^{\rm qs}  }{ 2 \times 10^3 } \right)^{-1}.
\label{eq:Tbin}
\end{eqnarray}
We expect that at least $\sim 4$ time bins are required to resolve a wave frequency, implying that the time frequency resolution is smaller than $\sim 1$~Hz. Thus, those detectors may not be capable of discerning normal temporal variations, unless the CCSN source is much closer than $10$~kpc or $\alpha$ is much higher than $0.2$. On the other hand, HK will register $\sim 7$ times the number of counts that SK will. Hence, temporal characteristics at $\sim 5$~Hz may then be resolvable.

As discussed in \citet{2021MNRAS.500..696N}, IceCube may have much better sensitivity with which to capture the temporal structure of supernova  neutrino signals by virtue of its large event count rate. Hence, let us make a similar estimation in this case. It should be noted that, unlike for other three detectors, the dominant component of detector noise is the detector itself for low event rates and Poisson noise for high event rates (and, hence, closer distances). More quantitatively, the noise can be estimated as \citep[see also][]{2011A&A...535A.109A,2013PhRvL.111l1104T,2021MNRAS.500..696N}:
\begin{eqnarray}
N^{\rm noise}_{\rm IC} (t) \sim \left( ( 1.48 \times 10^6 + n^{\rm qs} (t) ) \Delta {\rm T}_{\rm bin} \right)^{0.5}.
\label{eq:noiseIceCube}
\end{eqnarray}
$n^{\rm qs}$ is roughly 100 times higher than that in SK (it is $\sim 2 \times 10^5$ at this phase), indicating that the background noise dominates when the CCSNe is at a distance of $10$ kpc. Hence, the SNR can be given as
\begin{eqnarray}
{\rm SNR}_{\rm IC} \sim 10^{-3} \hspace{0.5mm} \alpha \hspace{0.5mm} n^{\rm qs} \hspace{0.5mm} \sqrt{\Delta {\rm T}_{\rm bin}}.
\label{eq:SNRatioIceCube}
\end{eqnarray}
Thus, the required time width to resolve the temporal variation can be estimated to be
\begin{eqnarray}
\Delta {\rm T}_{\rm bin (IC)} \sim 20 \hspace{0.5mm} [{\rm ms}] \left( \frac{ {\rm SNR} }{ 5 } \right)^2 \left( \frac{ \alpha }{ 0.2 } \right)^{-2}    \left( \frac{ n^{\rm qs} }{  2 \times 10^5 } \right)^{-2}.
\label{eq:TbinIC}
\end{eqnarray}
This estimate suggests that IceCube is capable of resolving temporal variations of $\sim 10$~Hz even when the CCSN is at $10$ kpc. Therefore, IceCube will provide the most detailed measurement among detectors of the temporal variations in the neutrino signal. It should be mentioned, however, that $\Delta {\rm T}$ for IceCube increases more rapidly with decreasing $n^{\rm qs}$ than that of other detectors (compare the $n^{\rm qs}$ dependence between Eqs.~\ref{eq:Tbin}~and~\ref{eq:TbinIC})\footnote{This is attributed to the fact that the background noise does not depend on $n^{\rm qs}$.}. This indicates that other detectors, in particular HK, would eventually become more sensitive to temporal variations at the very late phases ($\gtrsim 10$~s). It should also be mentioned that the threshold time (or frequency) bin strongly depends upon the distance to the CCSN. For instance, if the source is at $5$~kpc (the background noise still dominates in this situation), the threshold time frequency is more than $10$ times that at $10$~kpc, i.e., $\sim 100$~Hz temporal variations may be resolved. For such a nearby CCSN, IceCube would be capable of resolving $\sim 20$~Hz temporal variations even if $\alpha=0.1$, which corresponds to CCSN models with weak temporal variations such as the 16~$M_{\sun}$ model.

\begin{figure*}
  \rotatebox{0}{
    \begin{minipage}{1.0\hsize}
        \includegraphics[width=\columnwidth]{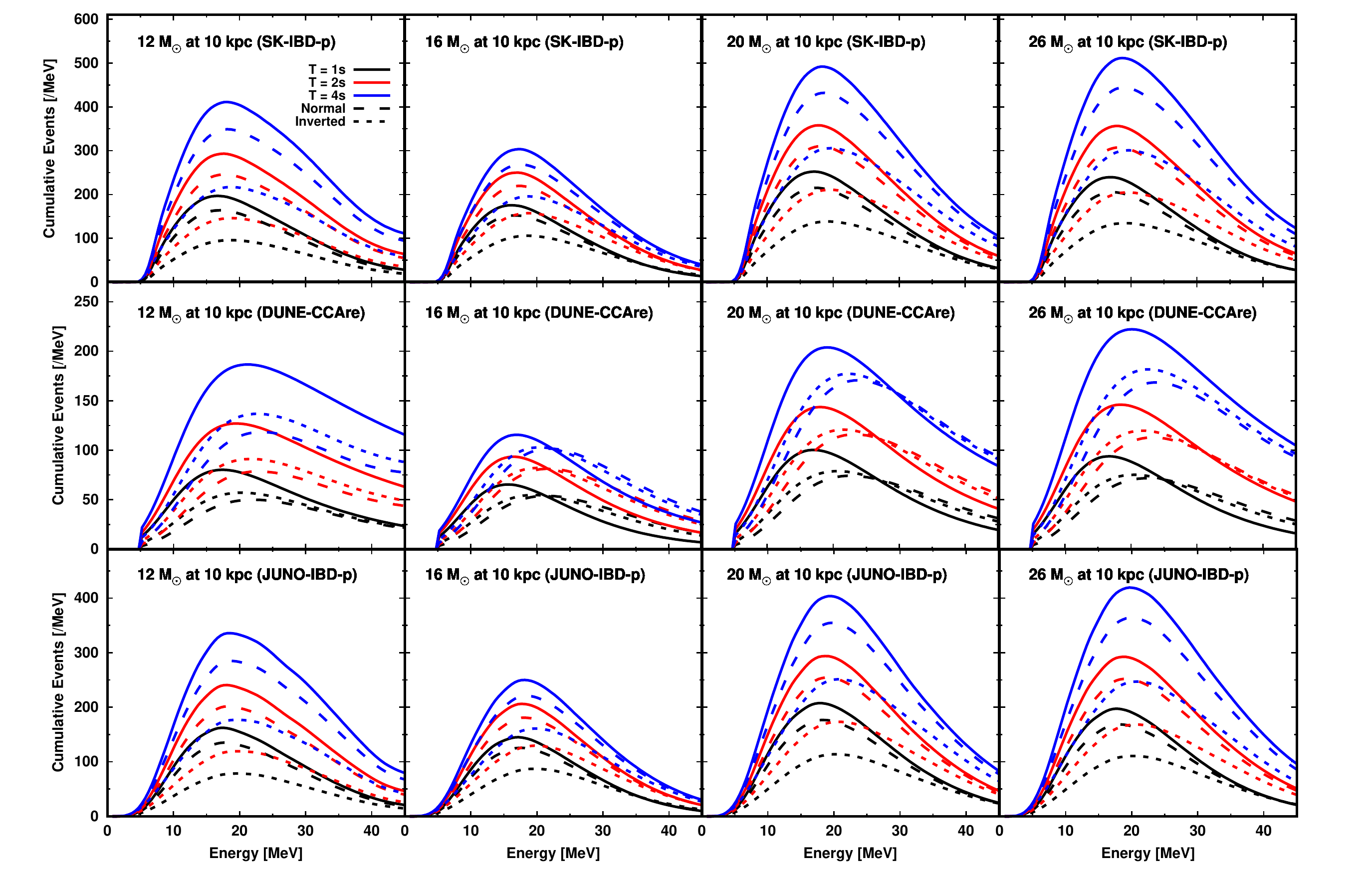}
    \caption{Energy spectrum of the cumulative number of events for the major reaction channel of each detector. We show the results for SK, DUNE, and JUNO from top to bottom. From left to right, we display the 12, 16, 20, and 26~$M_{\sun}$ models. The color represents the time. The line type distinguishes the neutrino oscillation models.}
    \label{graph_CumSpect_2Dlong}
  \end{minipage}}
\end{figure*}

We now turn our attention to properties of the cumulative number of events. Fig.~\ref{graph_CumSpect_2Dlong} shows the energy spectra with respect to each major reaction channel at each detector for selected post-bounce times: 1, 2, and 4 s. We note that the smearing effects of the detector response are taken into account in these plots (although Poisson noise is neglected). We find, as expected, that the spectral peak shifts with time to higher energies, regardless of the progenitor and neutrino oscillation model. It should be mentioned that the shape of the energy spectrum is similar for different CCSN models\footnote{There may be, however, a rich diversity of behaviors at higher energies ($\gtrsim 50$~MeV) among progenitors. See \citet{2020arXiv201015136N} for more details.}. 
On the other hand, the cumulative number of events in our CCSN models may be a bit higher than found by some.
This perception may due to slightly higher average energies (tens of percent) for one of our previous published F{\sc{ornax}} calculations \citep[see Fig.~4 in][]{2018JPhG...45j4001O}. Our Fig.~\ref{graph_neutrino_lumi_aveE_2Dlong} shows, however, that there is no such anomaly in our average neutrino energies compared with other multi-D CCSN models simulated by different groups \citep[see, e.g.,][]{2016ApJ...825....6S,2016ApJ...818..123B,2016ApJ...831...98R,2017MNRAS.472..491M,2018ApJ...855L...3O,2019ApJ...873...45G,2020ApJ...896..102K}, indicating that the neutrino signals presented in this paper would not deviate from the community norm. We note that that 1D model differs in the same way from what we have published in all our many multi-D papers. We have, however, not determined the reason for the slightly harder late-time spectrum of our 1D model in \citet{2018JPhG...45j4001O}. Therefore, we strongly encourage future community-wide efforts to quantify any differences among the different CCSN models. However, addressing this issue in the current paper in beyond its scope.

\begin{figure*}
  \rotatebox{0}{
    \begin{minipage}{0.8\hsize}
        \includegraphics[width=\columnwidth]{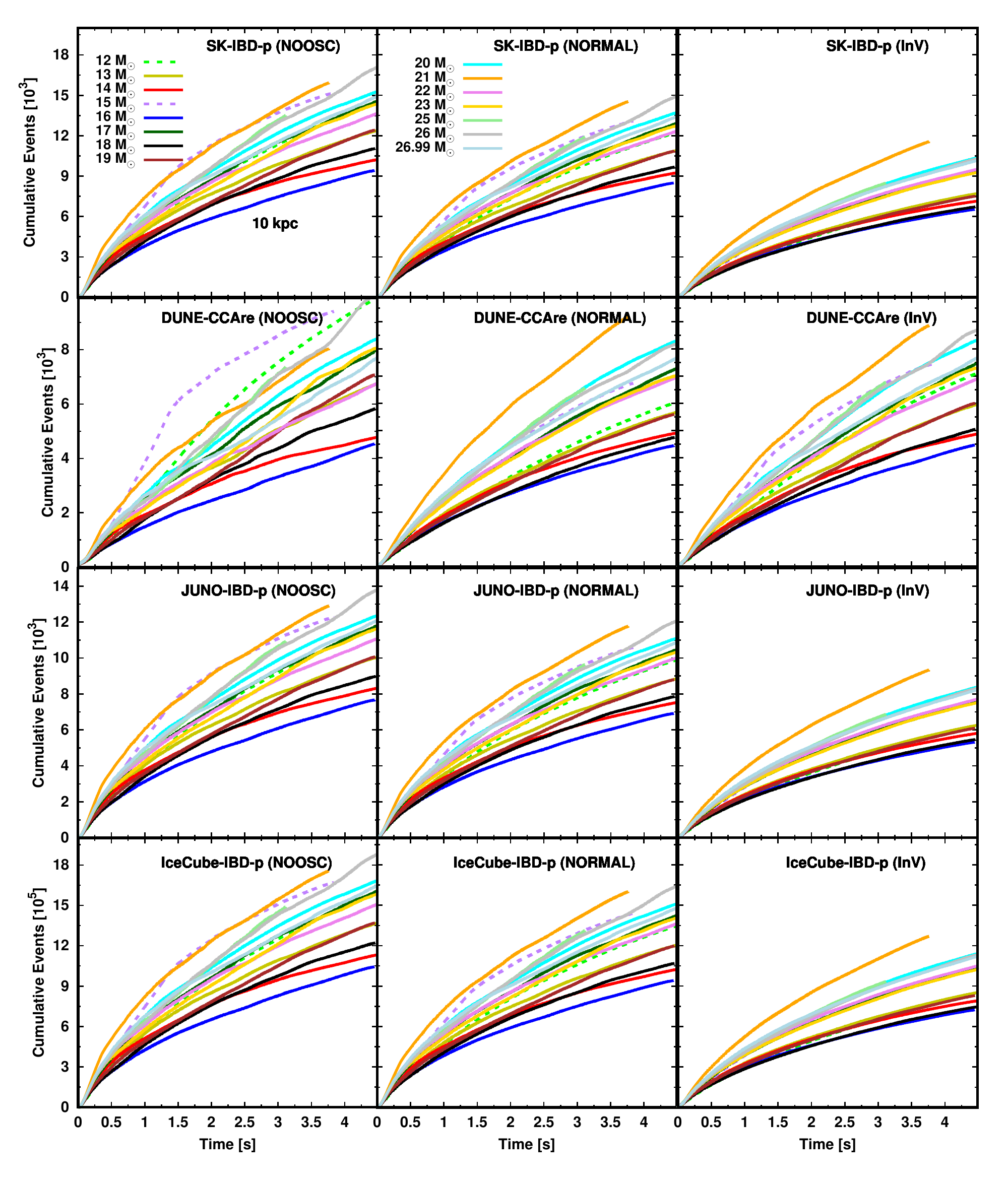}
    \caption{The time evolution of the cumulative number of events in the major reaction channel of each detector. Color and line type distinguishes CCSN models and explosion/non-explosion, respectively, which are the same convention as used in Fig.~\ref{graph_shockevo_2Dlong}. From top to bottom, we show the results of SK, DUNE, JUNO, and IceCube. From left to right, a different neutrino oscillation model is assumed.}
    \label{graph_Tevo_Cumu_2dlong}
  \end{minipage}}
\end{figure*}

Fig.~\ref{graph_Tevo_Cumu_2dlong} shows the time evolution of the cumulative number of events at each detector. In previous studies, it has been pointed out that the event count is positively correlated with the PNS mass \citep[see, e.g.,][]{2020arXiv200807070S}. We find, however, that such a correlation is not definitive and depends upon neutrino oscillation model and detector. For instance, the cumulative number of events at $\gtrsim 1$~s in DUNE for non-exploding models (12 and 15~$M_{\sun}$ models) tends to be higher than that of exploding models in the case without flavor conversion. We also note that the cumulative number of events at $\gtrsim 3$~s in the 26~$M_{\sun}$ model is the highest among exploding models, although the PNS mass is not the largest (see Fig.~\ref{graph_PNSmass_2Dlong}). Such a trend can be understood as follows. In common, these three models have high mass accretion rates at the late phases. This indicates that the accretion component of the $\nu_e$ and $\bar{\nu}_e$ emissions is also higher. However, the average neutrino energy of $\nu_e$s and $\bar{\nu}_e$s at the late phases is also higher than during the early phase. As a result, the detection efficiency of the neutrinos (in the case without flavor conversion) is higher. We note that the CCAre cross section in DUNE is more sensitive to high-energy neutrinos than is the IBD-p in SK, indicating that this average-energy difference makes more of a difference in DUNE than SK. Interestingly, the positive correlation between PNS mass and event rate in each detector tends to be recovered in the cases with flavor conversion (see, e.g., the case with the normal-mass hierarchy at DUNE in Fig.~\ref{graph_Tevo_Cumu_2dlong}). This can be attributed to the fact that all detectors have sensitivities to not only $\nu_e$s and $\bar{\nu}_e$s, but also to $\nu_x$s at the CCSN source by virtue of neutrino mixings. We note that the total $\nu_x$ count is strongly correlated with the mass of the PNS (see Fig.~\ref{graph_neutrino_lumi_aveE_2Dlong}).

Finally, we discuss the correlation between the cumulative number of events at each detector and the total neutrino energy (TONE) emitted by the supernova. This analysis is an extension of our previous discussion in Sec.~3.4 of \citet{2021MNRAS.500..696N}. In that previous study, we found an interesting correlation between cumulative event number and TONE, and provided fitting formulae for the relation. We note, however, that the formulae provided might be valid only in the early phase ($\sim 1$~s), since the event counts in the late phase were not available for the 3D models used there. Here, we update these fitting formulae by using the results of the present study.

\begin{figure*}
  \rotatebox{0}{
    \begin{minipage}{0.8\hsize}
        \includegraphics[width=\columnwidth]{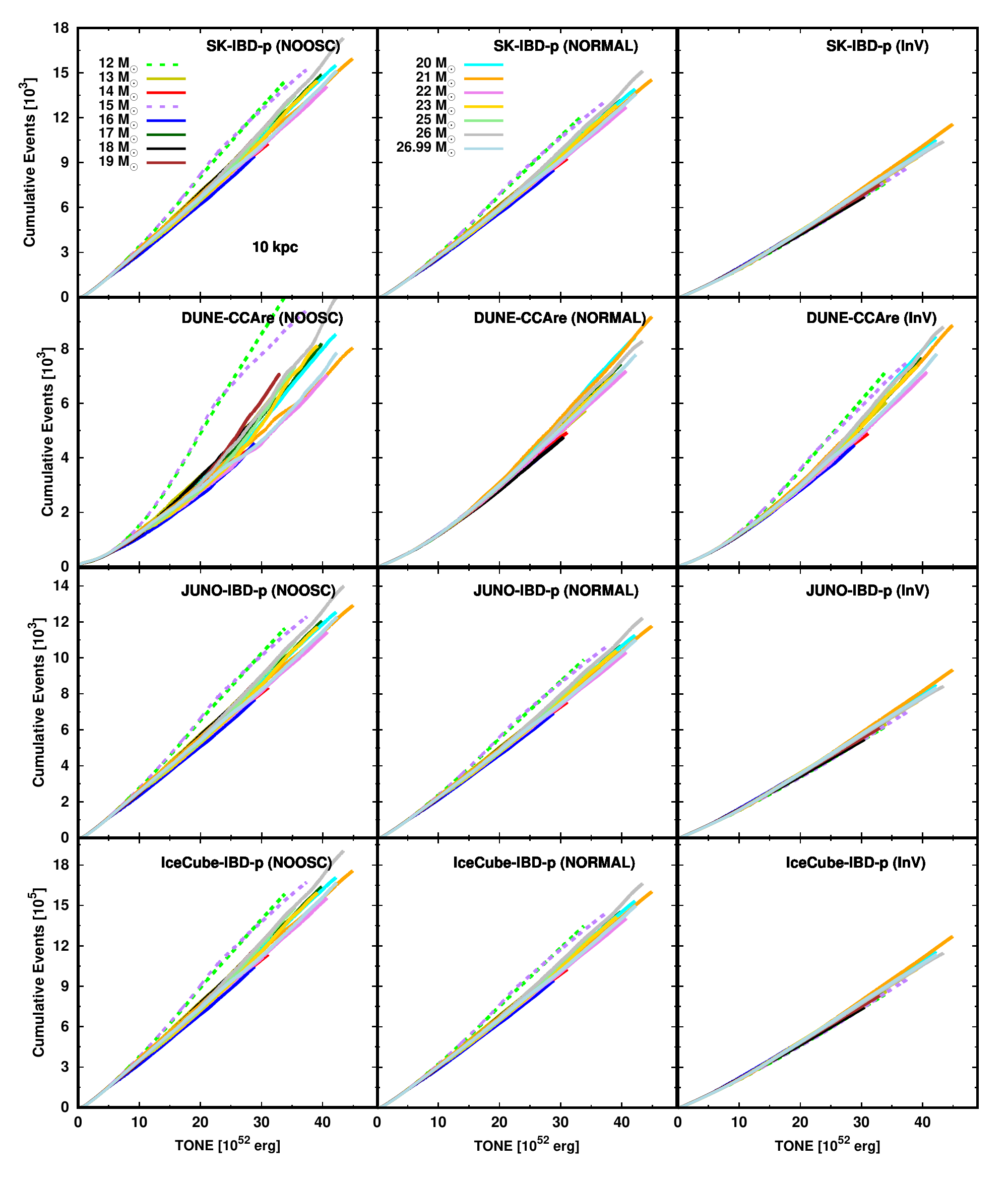}
    \caption{Cumulative number of events for the major reaction channel of each detector as a function of the total neutrino energy (TONE) emitted at CCSN sources ($10$~kpc). The position of each panel, color, and line type are the same as those in Fig.~\ref{graph_Tevo_Cumu_2dlong}.}
    \label{graph_EtotvsCum_2Dlong}
  \end{minipage}}
\end{figure*}

Fig.~\ref{graph_EtotvsCum_2Dlong} displays the cumulative number of events as a function of TONE at each detector with different neutrino oscillation models. First, we confirm that the fitting formulae presented in \citet{2021MNRAS.500..696N} recapitulate the relations very well in the early phase ($\sim 1$~s). In the late phase, however, largish deviations emerge in the non-exploding models (12 and 15~$M_{\sun}$) from the behavior of the explosion models in the case without flavor conversion (see the left column of Fig.~\ref{graph_EtotvsCum_2Dlong}). The systematic deviation of non-exploding models can be understood as follows: Non-exploding models manifest a high accretion component for $\nu_e$ and $\bar{\nu}_e$ emissions in the later phases, while the average energy is remarkably higher than during the early phases (see top-right and middle-right panels of Fig.~\ref{graph_neutrino_lumi_aveE_2Dlong}), which increases the detection efficiency for all detectors. As a result, the event counts tend to be higher with respect to the same TONE. On the other hand, the deviation is smaller in the cases with flavor conversions. For instance, an almost  progenitor-independent correlation emerges at DUNE for the normal-mass hierarchy. This is attributed to the fact that the event counts reflect $\nu_x$ at the CCSN source in the neutrino oscillation model. We note that $\nu_x$ constitutes the dominant contribution to TONE\footnote{We note that the neutrino luminosity of the individual species of heavy leptonic neutrinos is smaller than that of $\nu_e$ or $\bar{\nu}_e$ neutrinos. However, we have four such species.}. In the cases with other detectors (SK, HK, JUNO and IceCube), they also see a similar trend. It should be mentioned that for these detectors the progenitor dependence of the correlation is much smaller in the inverted-mass hierarchy than in the normal one, since $\bar{\nu}_e$ at the Earth mostly reflects the properties of the $\nu_x$ at the supernova.

Below, we provide approximate formulae for the correlations for the neutrino oscillation models. We first point out that the quadratic fit used in \citet{2021MNRAS.500..696N} can not capture the simulation results iat later times adequately. Hence, we fit them with a higher-order quartic polynominal. It should be noted that, although the fit can be improved by using cubic functions, we find that the functions break the monotonic relation before TONE reaches $6 \times 10^{53}$ erg. This is actually unphysical. Hence, we employ quartic functions in the fit. We confirm that monotonicity is guaranteed up to a TONE of $10^{54}$ erg, which is a firm upper limit to the total emission of CCSN neutrinos \citep[see also][]{2020PhRvD.102j3011R}.

The fitting formulae are given in the case of the normal mass hierarchy as:
\begin{eqnarray}
&&\hspace{-16.0mm} {\rm [SK-IBDp-NORMAL]} \nonumber \\
&&\hspace{-13.0mm} N_{\rm Cum} = \left( 220 \hspace{0.5mm} E_{52} + 5 \hspace{0.5mm} E_{52}^2 - 0.074 \hspace{0.5mm} E_{52}^3 + 0.0003 \hspace{0.5mm} E_{52}^4 \right) \nonumber \\
&&\left(  \frac{V}{32.5 \hspace{0.5mm} {\rm ktons}}  \right)
\left(  \frac{d}{10 \hspace{0.5mm} {\rm kpc}}  \right)^{-2}\, , 
\label{eq:fitSKNORMAL} \\
&&\hspace{-16.0mm} {\rm [DUNE-CCAre-NORMAL]} \nonumber \\
&&\hspace{-13.0mm} N_{\rm Cum} = \left( 90 \hspace{0.5mm} E_{52} + 4.5 \hspace{0.5mm} E_{52}^2 - 0.062 \hspace{0.5mm} E_{52}^3 + 0.00028 \hspace{0.5mm} E_{52}^4 \right) \nonumber \\
&&\left(  \frac{V}{40 \hspace{0.5mm} {\rm ktons}}  \right)
\left(  \frac{d}{10 \hspace{0.5mm} {\rm kpc}}  \right)^{-2}\, ,
\label{eq:fitDUNENORMAL} \\
&&\hspace{-16.0mm} {\rm [JUNO-IBDp-NORMAL]} \nonumber \\
&&\hspace{-13.0mm} N_{\rm Cum} = \left( 165 \hspace{0.5mm} E_{52} + 5.1 \hspace{0.5mm} E_{52}^2 - 0.082 \hspace{0.5mm} E_{52}^3 + 0.00039 \hspace{0.5mm} E_{52}^4  \right) \nonumber \\
&&\left(  \frac{V}{20 \hspace{0.5mm} {\rm ktons}}  \right)
\left(  \frac{d}{10 \hspace{0.5mm} {\rm kpc}}  \right)^{-2}\, ,
\label{eq:fitJUNONORMAL} \\
&&\hspace{-16.0mm} {\rm [IceCube-IBDp-NORMAL]} \nonumber \\
&&\hspace{-13.0mm} N_{\rm Cum} = \left( 23000 \hspace{0.5mm} E_{52} + 600 \hspace{0.5mm} E_{52}^2 - 9 \hspace{0.5mm} E_{52}^3 + 0.04 \hspace{0.5mm} E_{52}^4  \right) \nonumber \\
&&\left(  \frac{V}{3.5 \hspace{0.5mm} {\rm Mtons}}  \right)
\left(  \frac{d}{10 \hspace{0.5mm} {\rm kpc}}  \right)^{-2}\, ,
\label{eq:fitIceCubeNORMAL}
\end{eqnarray}
and in the case with the inverted mass hierarchy as
\begin{eqnarray}
&&\hspace{-16.0mm} {\rm [SK-IBDp-InV]} \nonumber \\
&&\hspace{-13.0mm}N_{\rm Cum} = \left( 170 \hspace{0.5mm} E_{52} + 4 \hspace{0.5mm} E_{52}^2 - 0.07 \hspace{0.5mm} E_{52}^3 + 0.00036 \hspace{0.5mm} E_{52}^4 \right) \nonumber \\
&&\left(  \frac{V}{32.5 \hspace{0.5mm} {\rm ktons}}  \right)
\left(  \frac{d}{10 \hspace{0.5mm} {\rm kpc}}  \right)^{-2}\, ,
\label{eq:fitSKInV} \\
&&\hspace{-16.0mm} {\rm [DUNE-CCAre-InV]} \nonumber \\
&&\hspace{-13.0mm}N_{\rm Cum} = \left( 90 \hspace{0.5mm} E_{52} + 4.5 \hspace{0.5mm} E_{52}^2 - 0.062 \hspace{0.5mm} E_{52}^3 + 0.00028 \hspace{0.5mm} E_{52}^4 \right) \nonumber \\
&&\left(  \frac{V}{40 \hspace{0.5mm} {\rm ktons}}  \right)
\left(  \frac{d}{10 \hspace{0.5mm} {\rm kpc}}  \right)^{-2}\, ,
\label{eq:fitDUNEInV} \\
&&\hspace{-16.0mm} {\rm [JUNO-IBDp-InV]} \nonumber \\
&&\hspace{-13.0mm}N_{\rm Cum} = \left( 135 \hspace{0.5mm} E_{52} + 3 \hspace{0.5mm} E_{52}^2 - 0.051 \hspace{0.5mm} E_{52}^3 + 0.0003 \hspace{0.5mm} E_{52}^4 \right) \nonumber \\
&&\left(  \frac{V}{20 \hspace{0.5mm} {\rm ktons}}  \right)
\left(  \frac{d}{10 \hspace{0.5mm} {\rm kpc}}  \right)^{-2}\, ,
\label{eq:fitJUNOInV} \\
&&\hspace{-16.0mm} {\rm [IceCube-IBDp-InV]} \nonumber \\
&&\hspace{-13.0mm}N_{\rm Cum} = \left( 18000 \hspace{0.5mm} E_{52} + 430 \hspace{0.5mm} E_{52}^2 - 7 \hspace{0.5mm} E_{52}^3 + 0.035 \hspace{0.5mm} E_{52}^4 \right) \nonumber \\
&&\left(  \frac{V}{3.5 \hspace{0.5mm} {\rm Mtons}}  \right)
\left(  \frac{d}{10 \hspace{0.5mm} {\rm kpc}}  \right)^{-2}\, ,
\label{eq:fitIceCubeInV}
\end{eqnarray}
where $N_{\rm Cum}$, $E_{52}$, and $V$ denote the cumulative number of events, TONE in the units of $10^{52} {\rm ergs}$, and the detector volume, respectively. We note that Eqs.~\ref{eq:fitSKNORMAL} and \ref{eq:fitSKInV} with $V=220$ ktons represent the HK case.

There are two caveats regarding the fitting formulae. First,
although they are capable of reproducing the results of explosion models, there is a systematic deviation for non-exploding models for all the detectors for the normal mass hierarchy, and for DUNE with the inverted mass hierarchy (see Fig.~\ref{graph_EtotvsCum_2Dlong}). This is attributed to the fact that the accretion component of $\nu_e$s or $\bar{\nu}_e$s (at the supernova) at late times contributes substantially to the event counts (as discussed already). As a result, the event counts tend to be higher than other cases with respect to the same TONE (see also Fig.~\ref{graph_Tevo_Cumu_2dlong} and relevant discussions). On the other hand, the systematic error is roughly $\sim 10 \%$, which is the same level of uncertainty due to the angular (observer direction) dependence \citep[see Secs.~3.2 and 3.4 in][]{2021MNRAS.500..696N}. This indicates that the errors may be overwhelmed by other uncertainties. We, hence, do not attempt any modifications to correct for the systematic deviations of non-exploding models. 
The cumulative number of events in our F{\sc{ornax}} CCSN models tends to be slightly higher than in others. This indicates that the TONE obtained by our fitting formulae could be underestimated.

The fitting formulae provided should be very useful in real observations, in particular for distant CCSNe. As discussed in \citet{2021MNRAS.500..696N,2021MNRAS.500..319N}, the TONE can be estimated through the retrieval of energy spectra for all flavors of neutrino by using purely observed quantities at multiple detectors. However, the statistical error is very large unless the CCSN source is very close and this implies that the retrieved TONE would not be accurate. Our fitting formulae, on the other hand, need only energy- and time-integrated (cumulative) event counts, which corresponds to the most statistically significant datum among observed quantities. For instance, the error for SK, JUNO, and DUNE for the Large Magellanic Cloud CCSNe ($\sim 50$~kpc) is $\lesssim 5 \%$, and HK will allow us to provide the TONE for CCSNe at the Andromeda galaxy ($\sim 700$~kpc) with $\sim 10\%$ errors. This indicates that the statistical noise does not compromise the accuracy of the estimation when compared with the estimate from the retrieved energy spectrum of all the flavors of neutrino \citep{2021MNRAS.500..319N}.

As an interesting demonstration, we apply our fitting formulae to estimate a TONE for SN 1987A from the event count in Kamiokande-II \citep{1987PhRvL..58.1490H}. We assume that all events were detected through the IBD-p reaction channel, and that the detector configuration is the same as that in SK except for the fiducial volume, which is $\sim 2$~ktons for Kamiokande-II. The cumulative number of events at Kamiokande-II was 11; our fitting formulae suggest that the TONE is $\sim 2 \times 10^{53}$~erg. By using the obtained TONE, we also estimate the mass of the neutron star in SN 1987A. For this, we assume that the TONE is the same as the binding energy of the NS. We also assume that the dimensionless tidal polarizability at $M=1.4~M_{\sun}$ ($\Lambda_{1.4}$) is $\sim 350$\footnote{This corresponds to the case using the SFHo EOS which we employ in our CCSN simulations \citep[see, e.g.,][]{2013ApJ...774...17S,2019PhRvD..99h3014H}. It is also within the observational constraints ($\Lambda_{1.4}=190^{+390}_{-120}$) placed by \citet{2018PhRvL.121p1101A}.}. By employing the result of \citep{2020PhRvD.102j3011R}, the gravitational mass of the neutron star can be estimated as $\sim 1.2~M_{\sun}$. The result seems consistent with that of other observed neutron stars \citep[see, e.g.,][]{2016ARA&A..54..401O}, albeit smaller than the canonical value ($\sim 1.4~M_{\sun}$). It should be noted that our estimation of TONE may be an underestimate (as mentioned earlier), which may account for the smaller estimated mass of the PNS in this example. The angular dependence of observer directions also alters the cumulative number by perhaps dozens of percent, affecting the estimate. One needs to keep in mind these uncertainties when we applying the fitting formulae to real observations.

\section{Summary and Conclusions}\label{sec:sumconc}

Our long-term 2D CCSN models for a wide range of progenitor mass reveal some new features in CCSN dynamics and concerning neutrino signals at the late phases. The matter dynamics in the vicinity of the PNS is highly variable, even during the later phases, due not only to PNS convection, but also to asymmetrical fallback mass accretion and fluid instabilities. It should be stressed that not excising the inner region of the CCSN core in a simulation is crucial for capturing all possible feedback effects on the neutrino signals. Using a self-consistent treatment throughout, we found that the temporal variations in the neutrino emissions mainly correlate with those in mass-accretion rate (see Fig.~\ref{graph_TimeCorrelation_NeutrinosMdot_shortTerm}). We also found that the correlation is generic for all explosion models, although the actual impact on the neutrino signals depends on model. We stress that the dynamical features in the neutrino signals during the late post-bounce phase are missing in  previous toy or spherically-symmetric models.

In this study, we  employed SNOwGLoBES, taking into account neutrino oscillations with an adiabatic model. We provided some basic results for the neutrino signals, such as the time evolution of the event rate (see Fig.~\ref{graph_evoEverate_2Dlong}) and energy spectra for the cumulative number of events (see Fig.~\ref{graph_CumSpect_2Dlong}) at each detector. We also assessed the detectability of event-rate temporal variations at each detector by employing a noise model (see Eqs.~\ref{eq:Tbin}~and~\ref{eq:TbinIC}). Not unexpectedly, we find that IceCube will be the best detector with which to study temporal variations. We have updated our fitting formulae for the correlation between cumulative number of events at each detector and the total neutrino energy (TONE) emitted at a CCSN source. Such formulae will prove very useful for low-statistic detections, i.e., distant CCSNe. Indeed, we present an interesting demonstration by using the real data for SN 1987A at Kamiokande-II, and we find the TONE is $\sim 2 \times 10^{53}$~erg and the corresponding (gravitational) NS mass could be near $\sim 1.2~M_{\sun}$. We note that once HK is available CCSNe at the Andromeda galaxy will also be targets.

There remain interesting issues to be addressed. It has been reported that stellar rotation affects the neutrino signal \citep[see, e.g.,][]{2018ApJ...852...28S} and we have yet to ascertain the degree to which our fitting formulae might be altered to accommodate it (Eqs.~\ref{eq:fitSKNORMAL}-\ref{eq:fitIceCubeInV}). It should be mentioned, however, that the effect should be minor, unless the rotation is remarkably faster than expected from stellar evolution and pulsar statistics. Another concern is with possible collective neutrino oscillations; indeed, there have been many reports that fast pairwise conversion could occur in both the preshock and post shock regions \citep{2019PhRvD.100d3004A,2019ApJ...886..139N,2020PhRvR...2a2046M,2020PhRvD.101b3018D,2020PhRvD.101f3001G,2020arXiv201206594A,2020arXiv201208525C}. Although this could have a significant impact on the neutrino signals in the early post-bounce phase ($ \lesssim 1$ s), it would have a minor effect in the late post-bounce phase, since the differences between the different neutrino spectra is then mild. Nevertheless, it will be important to pin down the magnitude of this correction. We are currently investigating those issues, and the results will be reported in forthcoming papers.

\section*{Acknowledgements}
The authors thank the anonymous referee for valuable comments. We acknowledge Kate Scholberg for help using the SNOwGLoBES software. We are also grateful for ongoing contributions to the effort of CCSN simulation projects by David Radice, Josh Dolence, Aaron Skinner, Matthew Coleman, and Chris White. We acknowledge support from the U.S. Department of Energy Office of Science and the Office of Advanced Scientific Computing Research via the Scientific Discovery through Advanced Computing (SciDAC4) program and Grant DE-SC0018297 (subaward 00009650). In addition, we gratefully acknowledge support from the U.S. NSF under Grants AST-1714267 and PHY-1804048 (the latter via the Max-Planck/Princeton Center (MPPC) for Plasma Physics). An award of computer time was provided  by the INCITE program. That research used resources of the Argonne Leadership Computing Facility, which is a DOE Office of Science User Facility supported under Contract DE-AC02-06CH11357. In addition, this overall research project is part of the Blue Waters sustained-petascale computing project, which is supported by the National Science Foundation (awards OCI-0725070 and ACI-1238993) and the state of Illinois. Blue Waters is a joint effort of the University of Illinois at Urbana-Champaign and its National Center for Supercomputing Applications. This general project is also part of the ``Three-Dimensional Simulations of Core-Collapse Supernovae" PRAC allocation support by the National Science Foundation (under award \#OAC-1809073). Moreover, access under the local award \#TG-AST170045 to the resource Stampede2 in the Extreme Science and Engineering Discovery Environment (XSEDE), which is supported by National Science Foundation grant number ACI-1548562, was crucial to the completion of this work. Finally, the authors employed computational resources provided by the TIGRESS high performance computer center at Princeton University, which is jointly supported by the Princeton Institute for Computational Science and Engineering (PICSciE) and the Princeton University Office of Information Technology, and acknowledge our continuing allocation at the National Energy Research Scientific Computing Center (NERSC), which is supported by the Office of Science of the US Department of Energy (DOE) under contract DE-AC03-76SF00098.

\section*{DATA AVAILABILITY}
The data underlying this article will be shared on reasonable request to the corresponding author.





\bibliographystyle{mnras}
\bibliography{bibfile}







\bsp	
\label{lastpage}
\end{document}